\documentclass[review]{elsarticle}
\usepackage{graphicx}
\graphicspath{{figures/}}

\usepackage[utf8]{inputenc}
\usepackage{subcaption}
\usepackage{algorithm}
\usepackage{algorithmic}
\usepackage{soul}

\usepackage{xcolor}
\usepackage{listings}

\definecolor{dkgreen}{rgb}{0,0.6,0}
\definecolor{gray}{rgb}{0.5,0.5,0.5}
\definecolor{mauve}{rgb}{0.58,0,0.82}

\journal{Journal of Simulation Modelling Practice and Theory}

\begin{document}

\begin{frontmatter}

\title{A Unified Cloud-Enabled Discrete Event Parallel and Distributed Simulation Architecture}

\author[ucm]{José L. Risco-Martín\corref{cor1}}
\ead{jlrisco@ucm.es}
\author[ucm]{Kevin Henares}
\ead{khenares@ucm.es}
\author[mitre]{Saurabh Mittal}
\ead{smittal@mitre.org}
\author[ucm]{Luis F. Almendras}
\ead{luisalme@ucm.es}
\author[ucm]{Katzalin Olcoz}
\ead{katzalin@ucm.es}

\cortext[cor1]{Corresponding author}
\address[ucm]{Dept. of Computer Architecture and Automation, Universidad Complutense de Madrid, Madrid 28040, Spain}
\address[mitre]{The MITRE Corporation, Fairborn, OH, USA}

\begin{abstract}
Cloud infrastructure provides rapid resource provision for on-demand computational requirements. Cloud simulation environments today are largely employed to model and simulate complex systems for remote accessibility and variable capacity requirements. In this regard, scalability issues in Modeling and Simulation (M\&S) computational requirements can be tackled through the elasticity of on-demand Cloud deployment. However, implementing a high performance cloud M\&S framework following these elastic principles is not a trivial task as parallelizing and distributing existing architectures is challenging. Indeed, both the parallel and distributed M\&S developments have evolved following separate ways. Parallel solutions has always been focused on ad-hoc solutions, while distributed approaches, on the other hand, have led to the definition of standard distributed frameworks like the High Level Architecture (HLA) or influenced the use of distributed technologies like the Message Passing Interface (MPI). Only a few developments have been able to evolve with the current resilience of computing hardware resources deployment, largely focused on the implementation of Simulation as a Service (SaaS), albeit independently of the parallel ad-hoc methods branch. In this paper, we present a unified parallel and distributed M\&S architecture with enough flexibility to deploy parallel and distributed simulations in the Cloud with a low effort, without modifying the underlying model source code, and reaching important speedups against the sequential simulation, especially in the parallel implementation. Our framework is based on the Discrete Event System Specification (DEVS) formalism. The performance of the parallel and distributed framework is tested using the xDEVS M\&S tool, Application Programming Interface (API) and the DEVStone benchmark with up to eight computing nodes, obtaining maximum speedups of $15.95\times$ and $1.84\times$, respectively.

\end{abstract}


\end{frontmatter}

\section{Introduction}\label{sec:intro}

Parallel and distributed simulation fields are two distinct fields that emerged in the 1970s and 1980s respectively from two different research communities \cite{Fujimoto2000}. The Parallel Simulation community was focused on accelerating simulations through the exploitation of high-performance computing (HPC) resources. Accordingly, the parallel simulation is defined as the parallelizing of simulation across different computing nodes. When there is a significant geographical separation between the computing nodes, a parallel simulation turns into a distributed simulation. While the parallel computing solution is implicitly distributed, the converse is not always true. The Distributed Simulation community (independent of the parallel simulation aspect) has largely focused on interconnecting partial simulations through local or wide area networks. Currently, theses two communities continue to keep the same driving force: parallel simulation works mainly over tightly coupled hardware entities, while distributed simulation still works on loosely coupled components communicating over standards-based wide area networks (e.g., Distributed Interactive Simulation [DIS], High level Architecture [HLA], etc.). 

The desire to bring both parallel and distributed M\&S faces new challenges, due to the complexity of new applications and the evolution in the underlying hardware \cite{Fujimoto2016}. From an application point of view, simulating systems of ever increasing complexity such as those in Internet of Things, needs huge computational power \cite{Dangelo2017}. 

On the other hand, from the hardware point of view, new paradigms such as Cloud Computing enables the provision of the large computational power of Google or Amazon infrastructure to a single researcher to exploit the computing resources for simulation execution \cite{Kratzke}. However, the technologies for Cloud computing require specific handling and the M\&S applications need to evolve to adapt to cloud-enabled architectures \cite{manifesto}.

Parallel and distributed simulation in the cloud is an emerging research area driven by the cost advantages of scaling simulations with available on-demand computing resources, without incurring the expense of purchasing and operating high-performance computing platforms, an issue that has prevented the adoption of parallel and distributed simulation technology in the past \cite{Kewley2016}. According to the U.S. National Institute of Standards and Technology (NIST), \emph{Cloud computing is a model for enabling ubiquitous, convenient, on-demand network access to a shared pool of configurable computing resources (e.g., networks, servers, storage, applications, and services) that can be rapidly provisioned and released with minimal management effort or service provider interaction} \cite{NIST}. Given this heterogeneity, implementing the appropriate simulation computational infrastructure is a very sophisticated task. 

Previous M\&S engines have been designed mainly to tackle global challenges like transparency, simulation as a service, cost, and performance \cite{Dangelo2014}. The solution presented in this paper exploits the categorical separation of modeling and simulation aspect in any M\&S architecture and focuses on the same model that is executed through parallel or distributed simulation execution, i.e., given a standard model, it must be simulated in a  sequential, parallel or distributed contexts without changing a single line of the model source code. To achieve this goal:

\begin{enumerate}
    \item The model must be defined following standard specifications,
    \item Simulation engine and model must be decoupled, and
    \item The simulation technology must be resilient enough to easily address the computing diversification offered by the Cloud: virtualization, containerization, etc.
\end{enumerate}

There exist several M\&S formalisms that help to deal with \#1 above. Among them, we have selected the Discrete Event System Specification (DEVS) \cite{Zeigler2000}, since it provides not only a global framework to define models, but also standard mechanisms to develop the simulation engine, which is categorically decoupled from the model, addressing also point \#2. 

Although the parallel solution is always easy to deploy, the distributed approaches present many technical difficulties when deploying a distributed simulation through a cluster, a set of virtual machines or containers, etc. To facilitate a better distribution mechanism, our solution makes use of a straightforward distributed architecture, a client/server pattern using standard sockets. We present a unified event-driven parallel and distributed simulation architecture, where sequential simulations can be scaled-up to a parallel and distributed simulation execution with an extremely easy deployment mechanism, even in a cloud-enabled environment.

The main contributions of our research can be summarized as follows:
\begin{itemize}
    \item We propose a unified parallel and distributed simulation architecture using the DEVS formalism as implemented through xDEVS Tool and API \cite{github2014xdevs}. Once the model has been implemented, it can be simulated in sequential, parallel or distributed platforms, without modifying a single line of the model's source code.
    \item A simulation deployment standard scheme is designed, where using standard XML files supported by schema definitions, the simulation can be deployed in parallel or distributed platforms.
    \item The multi-modal deployment is highly resilient, i.e., it can be done in several centralized or cloud-based resources, including physical (real) or virtual machines, or containers using the proposed method and structure.
    \item The standard DEVStone benchmark is revisited to consider (for the first time) the best ways of evaluating parallel and distributed DEVS-based simulations.
    \item The evaluation of the proposed architecture is done not only considering the traditional performance metric, but also resources distribution and cost.
\end{itemize}

This paper is organized as follows. Related work is described in Section \ref{sec:related}. Section \ref{sec:technologies} introduces the foundational technologies behind the work developed through this research. In Section \ref{sec:architecture}, a detailed view of the parallel and distributed architecture implemented in xDEVS is presented. Section \ref{sec:deployment} shows the parallel and distributed deployment options. Both parallel and distributed approaches are configured and evaluated in Section \ref{sec:evaluation}. Finally, we present conclusions and future work in Section~\ref{sec:conclusion}.

\section{Related work}\label{sec:related}

A plethora of parallel and distributed simulation architectures can be found in literature. The parallel simulation paradigm has usually brought ad-hoc solutions using multi-threaded programming technologies \cite{Bagrodia1998} \cite{Aniszewski2021}, although more specific solutions can be found in the last decade using Graphic Processing Units (GPUs) \cite{Ubal2012} or even Field Programmable Gate Arrays (FPGAs) \cite{Wang2018}. On the other hand, the distributed simulation paradigm has driven the development of not only distributed technologies like Message Passing Interface (MPI) \cite{Pelkey2011}, but also general and robust distributed simulation standards such as the DIS \cite{Hofer1995} or the HLA \cite{Dahmann1997}.

With respect to the DEVS M\&S formalism used in this work, many simulation engines have been developed and published during the last twenty years. Some of them have been specially designed to handle parallel or distributed simulations. 

Regarding parallel DEVS implementations, we may find the works of Liu  \cite{Liu2009}, Nutaro \cite{Nutaro2009}, or Lanuza \cite{Lanuza2020} among others. These developments are based on optimistic simulators following the concept of Logical Processes, Time Warp Algorithm, or others. However, these approaches do not provide standard interfaces to facilitate performance evaluation and comparison through parallel or distributed benchmarks. The development of xDEVS M\&S engine dates back to our first text on DEVS Unified Process \cite{MittalMartin2013} and over the years has been extended to bring in various domain specific languages (DSLs) to the DEVS world \cite{C-Mittal2016}. Our earlier approaches \cite{risco2008optimization}  parallelize the standard DEVS simulation loops that call transition and output functions, maintaining the original DEVS specification and all its properties. The codebase is currently maintained at \cite{github2014xdevs}.

With respect to the distributed  DEVS implementations, some frameworks like DEVS/SOA \cite{J-Mittal2009a}, or CD++ \cite{AlZoubi2009}, currently deprecated, were based on the concept of Simulation as a Service (SaaS), while others like PyPDEVS \cite{Tendeloo2015} that are more flexible require the user to be aware of intricacies to distribute the simulation.

\section{Foundational technologies}\label{sec:technologies}
Our framework must be able to execute simulations and optimization studies in distributed and decentralized environments. To this end, we have selected a container-based distributed architecture based on microservices due to its potential and configuration simplicity \cite{Gannon}. In this section, we describe the technologies involved that perform distributed simulations based on microservices, containerization paradigms and DEVS formalism.

\subsection{Microservices and containerization paradigms}

Traditionally, systems have been developed following monolith architectures, where the entire system's function is based on a single program. This monolith model often results in tightly-coupled systems, with highly interconnected and interdependent components.

In contrast, microservices architectures have been gaining traction and popularity over the last few years. In these architectures, the different features of a system are decomposed into separated application units, which communicate with each other primarily through asynchronous event-driven mechanisms. A standard communication protocol and a set of well-defined APIs independent of any vendor, product, or technology are used for inter-microservice communications. As Mittal and Martin \cite{C-Mittal2017} point out, any microservices-based architecture has to address two fundamental issues: distributed data management (to store the state of the microservice locally) and shared event processing (to facilitate the information exchange between stateless microservices). This information from the local data and the event processing is kept inside the microservices and is used together to execute their inherent business logic. This alternative methodology results in (i) the development of more resilient systems, as the system continue its operation even if specific components go down, (ii) better use of the resources, as it allows to scale specific components based on the demand, (iii) clear independence of the system's components, that can be developed and tested separately.

\begin{figure}[t]
\centering
\includegraphics[width=0.67\textwidth]{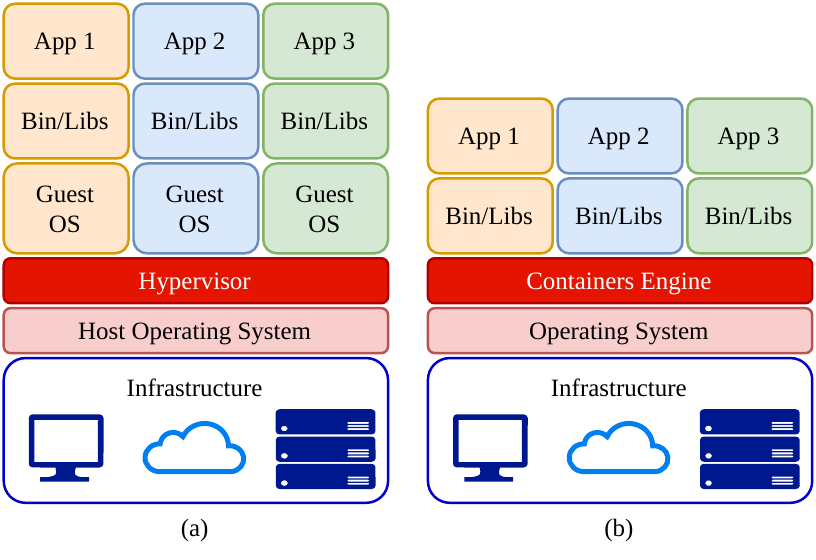}
\caption{Graphical representation of (a) virtual machine and (b) container architectures}\label{fig:vm_vs_docker}
\end{figure}

To implement and deploy microservices-based systems, it is customary to use a containerized architecture. A container is a lightweight, efficient, and standard way for applications to move between environments and run independently. It wraps a piece of software in a complete file system that contains everything needed to run (except for the shared operating system on the server). This approach favors the portability of systems, as they can be easily deployed in a multitude of operating systems and hardware architectures, and allows to accelerate development, test, and production cycles.  
They also present less overhead than traditional virtual machine environments, as they do not include operating system images. As a result, in many cases, the traditional virtualization present in the first times of Cloud Computing is transitioning towards container-based architectures. Fig.~\ref{fig:vm_vs_docker} illustrates the differences between these two approaches. In particular, Fig.~\ref{fig:vm_vs_docker}.a shows how virtual machines store the whole Operating System (OS), libraries, binaries, and applications, requiring a huge memory space in the host machine. Fig.~\ref{fig:vm_vs_docker}.b shows how a container is composed by libraries, required binaries, and applications; and how all the containers share the same OS kernel.

When managing large container-based systems, container orchestration becomes essential. This orchestration is in charge of automating the deployment, management, scaling, networking, and availability of the containers. As these practices became established, different tools emerged that encapsulate them and allow them to be applied in different container engines. Some popular examples of these container orchestration tools are Kubernetes and Docker Swarm. Moreover, many cloud services offer Infrastructure as a Service (IaaS) platforms based on these tools allowing developers to deploy complex container-based scenarios. Among them are Amazon Elastic Kubernetes Service (EKS), Azure Kubernetes Service (AKS), and Google Kubernetes Engine (GKE).

\subsection{Discrete Event System (DEVS) specifications}

DEVS is a general formalism for discrete event systems modeling based on mathematical Set theory~\citep{Zeigler2000}. We can distinguish between Classic DEVS and Parallel DEVS. Parallel DEVS was introduced as a revision of Classic DEVS. Moving forward, any mention of DEVS implies Parallel DEVS. The notion of parallelism in DEVS formalism exists at both the modeling and simulation layers. It is the confluence of events that happen concurrently at a given instant and how the DEVS formalism handles this confluence of events in its model specification and eventually implements it in the simulation coordinators preserving this confluence. The DEVS formalism does not address the performance aspect of parallel computing for speedup, etc. The execution of DEVS coordinator and component simulators in a multi-core architecture is one of the topics explored in this paper and is described ahead. 

The DEVS formalism includes two model types: atomic and coupled models. Both models have an interface consisting of input ($X$) and output ($Y$) ports to communicate with others. In atomic models, every state ($Q$) in the model is associated with the time advance function $ta$, which determines the duration during which the state remains unchanged. Once the time assigned to the state has passed, an internal transition function ($\delta_{\rm int}: Q \leftarrow Q$) is fired and an internal transition is triggered, producing a local state change ($\delta_{\rm int}(s) = s'$). At that moment, the model execution results are spread through the model’s output ports by activating an output function ($\lambda$). Input external events (events received from other models) are collected in the input ports. An external transition function ($\delta_{\rm ext}: Q \times X \leftarrow Q$) specifies how to react to those inputs, using the current state ($S$), the elapsed time since the last event ($e$) and the input value ($X$) ($\delta_{\rm int}((s, e), x) = s'$). Parallel DEVS introduces a confluent function ($\delta_{\rm int}((s, ta(s)), x) = s'$), which decides the next state in cases of collision between external and internal events. 

A coupled model has four additional sets: children components $C$, the external input $EIC$, external output $EOC$, and internal coupling $IC$ relations. Coupled models represent the aggregation/composition of two or more atomic and coupled models connected by explicit couplings, making DEVS closed under coupling. Closure under coupling allows to use networks of systems as components in a larger coupled systems, leading to hierarchical, modular construction. Overall, this formalism provides a framework for information modeling that has several advantages to analyze and design complex systems: completeness, verifiability, extensibility and maintainability.

\section{Parallel and distributed architecture}\label{sec:architecture}

Once a system is described according to DEVS theory, it can be easily implemented using one of the many DEVS M\&S engines. They all offer a programmer-friendly API to define new models using a high level language, but only a few provide a user-friendly API for parallel and distributed simulation execution. Among them, xDEVS \cite{github2014xdevs, J-RiscoMartin2017, C-Mittal2017} has recently incorporated a good alternative to parallelize and distribute simulations in the Cloud, following the microservices architecture and containerization mentioned in the previous section. This section provides a brief introduction to xDEVS, followed by both the parallel and distributed architectures.

\begin{figure}
\centering
\includegraphics[width=0.9\textwidth]{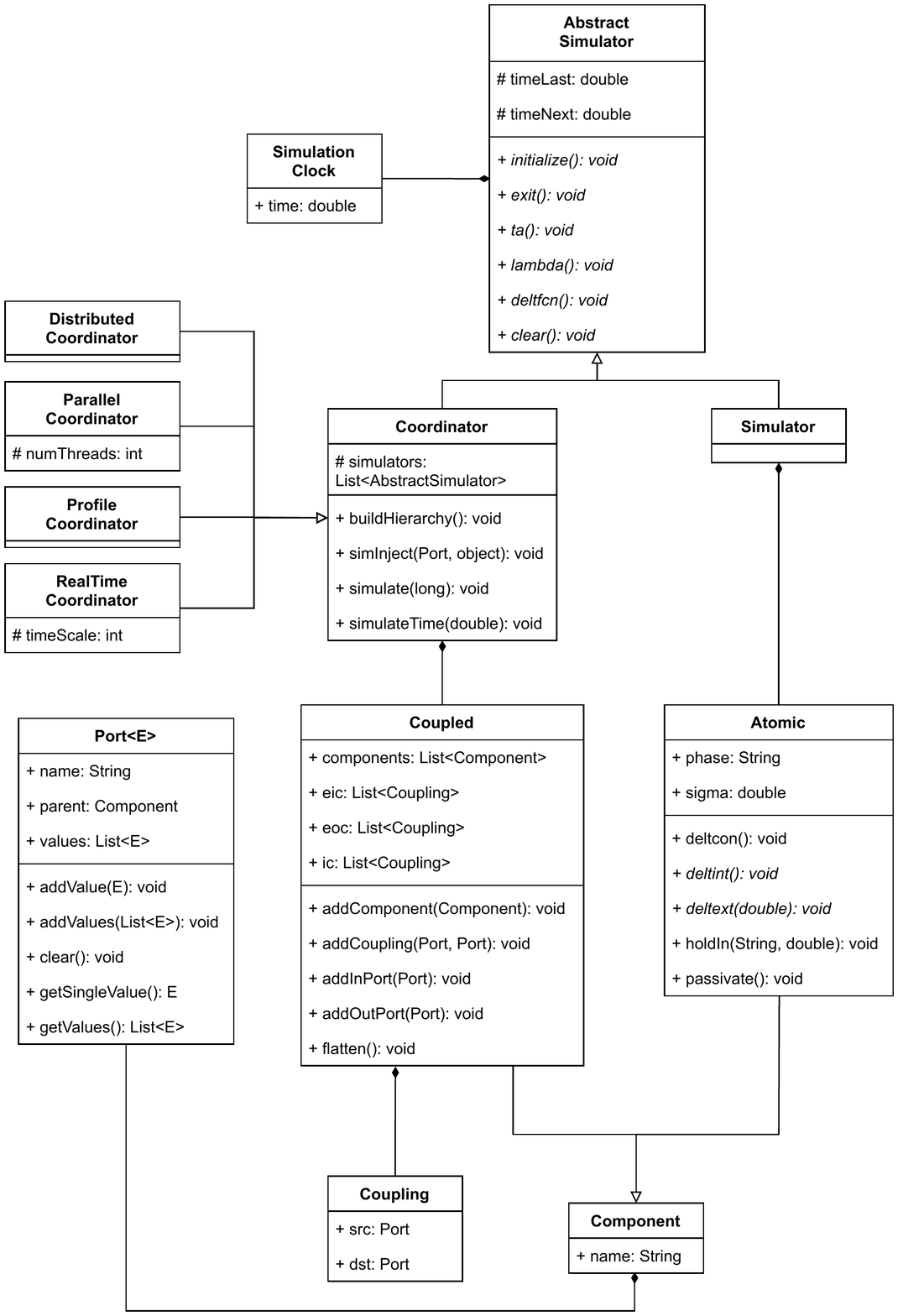}
\caption{Class diagram of the xDEVS architecture.}
\label{fig:xdevs_class_diagram}
\end{figure}

\subsection{xDEVS}

xDEVS is a cross-platform discrete event system simulator that provides a universal DEVS Application Programming Interface (API) both at the modeling and the simulation levels. The API is realized in three widely used object-oriented programming languages: C++, Java, and Python. The repository is made available through an API project at \cite{github2014xdevs}, where the project has three principal branches (named c++, java, and python). This framework allows the specification and execution of DEVS models. Based on the DEVS formalism, it has a clear separation between the modeling and simulation layers. A class diagram showing the relationship between these modeling and simulation layers is shown in Figure~\ref{fig:xdevs_class_diagram}.

DEVS models in xDEVS are created using two main components. \texttt{Atomic} components define the behavior of the system. \texttt{Coupled} components contains other  \texttt{Atomic} and \texttt{Coupled} components, creating a model hierarchy. Both of them have \texttt{Ports}, that represent input/output information points. To link two components of the model a \texttt{Coupling} can be created, selecting the source and destination \texttt{Ports}. The information of \texttt{Couplings} is contained in the \texttt{Coupled} elements that wrap the ports to be linked.

The simulation layer is based on the concept of the Abstract Simulator. Following this concept we divide the simulation entities in \texttt{Simulators} and \texttt{Coordinators}. Each \texttt{Simulator} is related to an \texttt{Atomic} component. Each \texttt{Coordinator} is attached to a specific coupled model and synchronize their child \texttt{Simulators} and \texttt{Coordinators}. This results in an equivalent hierarchy to the one described for the modeling layer.

Accordingly, the Coordinator API deals with executing a DEVS coupled model over time. In this paper, we present both the parallel and distributed coordinators, named \texttt{CoordinatorParallel} and \texttt{CoordinatorDistributed}, recently designed to allow simulations in centralized or distributed parallel computing environments. 

\subsection{Parallel architecture}

The xDEVS parallel coordinator executes the sequential coordinator using multiple concurrent threads and is apt for multi-core machines with a shared memory subsystem. An xDEVS parallel coordinator is formed by several thread pools. Each coordinator child, generally a simulator\footnote{By default, the root coupled model is flattened in parallel and distributed simulations. A flattened DEVS model is a model that is reduced to a single level coupled model containing only atomic models as a result of a flattening algorithm that preseves the coupling relationships. As a consequence, the root coordinator only manages simulators (for atomic components) and no hierarchial coordinators. This behavior can be changed by the modeler.}, is attached to one of the thread pools.

Listing \ref{lst:parallel} shows a code excerpt of a parallel coordinator with a single thread pool. As can be seen when building the hierarchy, a couple of tasks, instances of \texttt{TaskDeltFcn} and \texttt{TaskLambda}, are created for each simulator: one task to run the transition function and another one to run the output function, respectively. 

\begin{lstlisting}[language=java,label={lst:parallel},caption={xDEVS parallel coordinator},captionpos=b]]
public class CoordinatorParallel extends Coordinator {
  protected int numberOfThreads;
  protected LinkedList<TaskLambda> lambdaTasks = new LinkedList<>();
  protected LinkedList<TaskDeltFcn> deltfcnTasks = new LinkedList<>();
  protected ExecutorService executor;
  public CoordinatorParallel(SimulationClock clock, Coupled model, int numberOfThreads) {
    super(clock, model, true);
    this.numberOfThreads = numberOfThreads;
    executor = Executors.newFixedThreadPool(numberOfThreads);
  }
  public void buildHierarchy() {
    super.buildHierarchy();
    simulators.forEach((simulator) -> {
      lambdaTasks.add(new TaskLambda(simulator));
    });
    simulators.forEach((simulator) -> {
      deltfcnTasks.add(new TaskDeltFcn(simulator));
    });        
  }
  public void lambda() {
    executor.invokeAll(lambdaTasks);
    propagateOutput();
  }
  public void deltfcn() {
    propagateInput();
    executor.invokeAll(deltfcnTasks);
    tL = clock.getTime();
    tN = tL + ta();
  }
}
\end{lstlisting}

The DEVS simulation loop basically consists of executing in all the simulators the following:
\begin{enumerate}
    \item the time advance function,
    \item the output function, and
    \item the transition function.
\end{enumerate}
The time advance function invokes each simulator for the next time event, so it is not parallelized because of low complexity. Output and transition functions, on the contrary, can require more CPU time. Thus, these two tasks are fully parallelized in the thread pool. As Listing \ref{lst:parallel} shows, both the output and transition functions run the corresponding child functions in parallel (through the \texttt{invokeAll} call), with a number of threads defined by the user (in the attribute \texttt{numberOfThreads}). 

The modeler can add more thread pools by creating a new parallel coordinator with two or more \texttt{ExecutorService} thread pools. Then, both the lambda and transition functions must be modified following this schema for $N$ pools (Listing \ref{lst:poolschema}):

\begin{lstlisting}[language=Java,label={lst:poolschema},caption={Schema for N pools},captionpos=b]
  // ...
  public void lambda() {
    executor1.invokeAll(lamdaTasks1);
    executor2.invokeAll(lamdaTasks2);
    // ...
    executorN.invokeAll(lamdaTasksN);
  } 
\end{lstlisting}

It is worthwhile to mention that different pools are executed sequentially, although each one internally is run in parallel. However, having a big thread pool (with many threads) for complex models and a small pool (with a few threads) for lighter models can be interesting in some cases, subject to further investigation.

For the purposes of this research paper, we have created a simple and specific class that loads an XML file, which defines the allocation pool for each atomic model. Thus, the class creates as many different threads pools as those defined in the XML file along with the number of threads for each pool.

Note that this parallelization is completely DEVS compliant, since it follows the DEVS simulation algorithm defined in \cite{Zeigler2000}. Thus, we can assure that the results of the parallel simulation will be equivalent, and indeed identical, to those obtained with the sequential simulation. Actually, the same \texttt{Couple} model can be simulated with the sequential \texttt{Coordinator} class and the parallel \texttt{CoordinatorParallel} class, without changing a single line in the model source code.

\subsection{Distributed architecture}

In the following we provide the details of the design and implementation of the xDEVS distributed simulation engine. Its novelty and strength resides in simplifying the approaches developed during the last decade to ease the deployment of DEVS-based distributed simulations, agnostic of the heterogeneity of the Cloud solution in use. 

\subsubsection{Overview}

\begin{figure}
    \centering
    \begin{subfigure}{0.45\textwidth}
        \includegraphics[width=\textwidth]{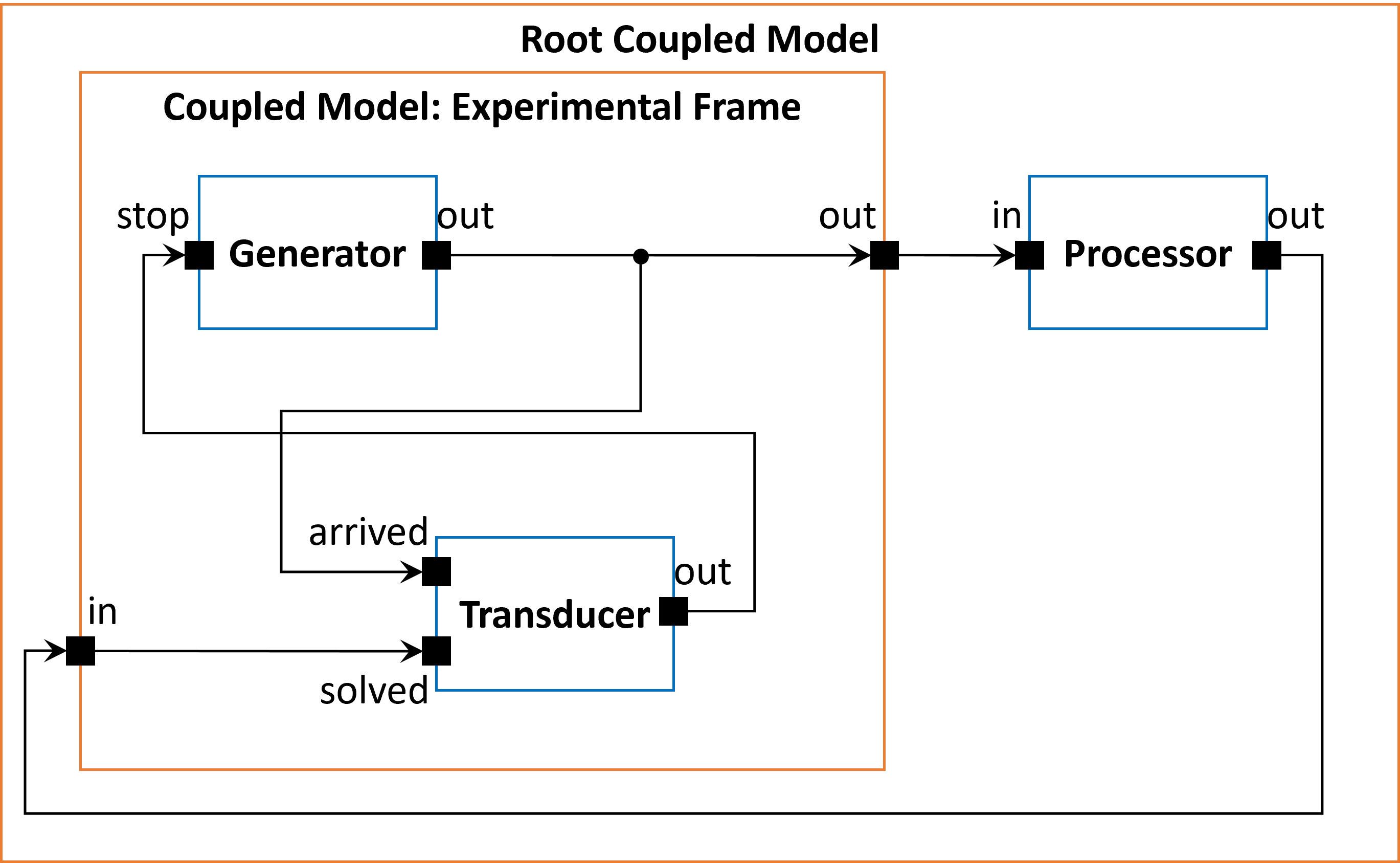}
        \caption{DEVS hierarchical coupled model}
        \label{fig:efp}
    \end{subfigure}\hfill
    \begin{subfigure}{0.45\textwidth}
        \includegraphics[width=\textwidth]{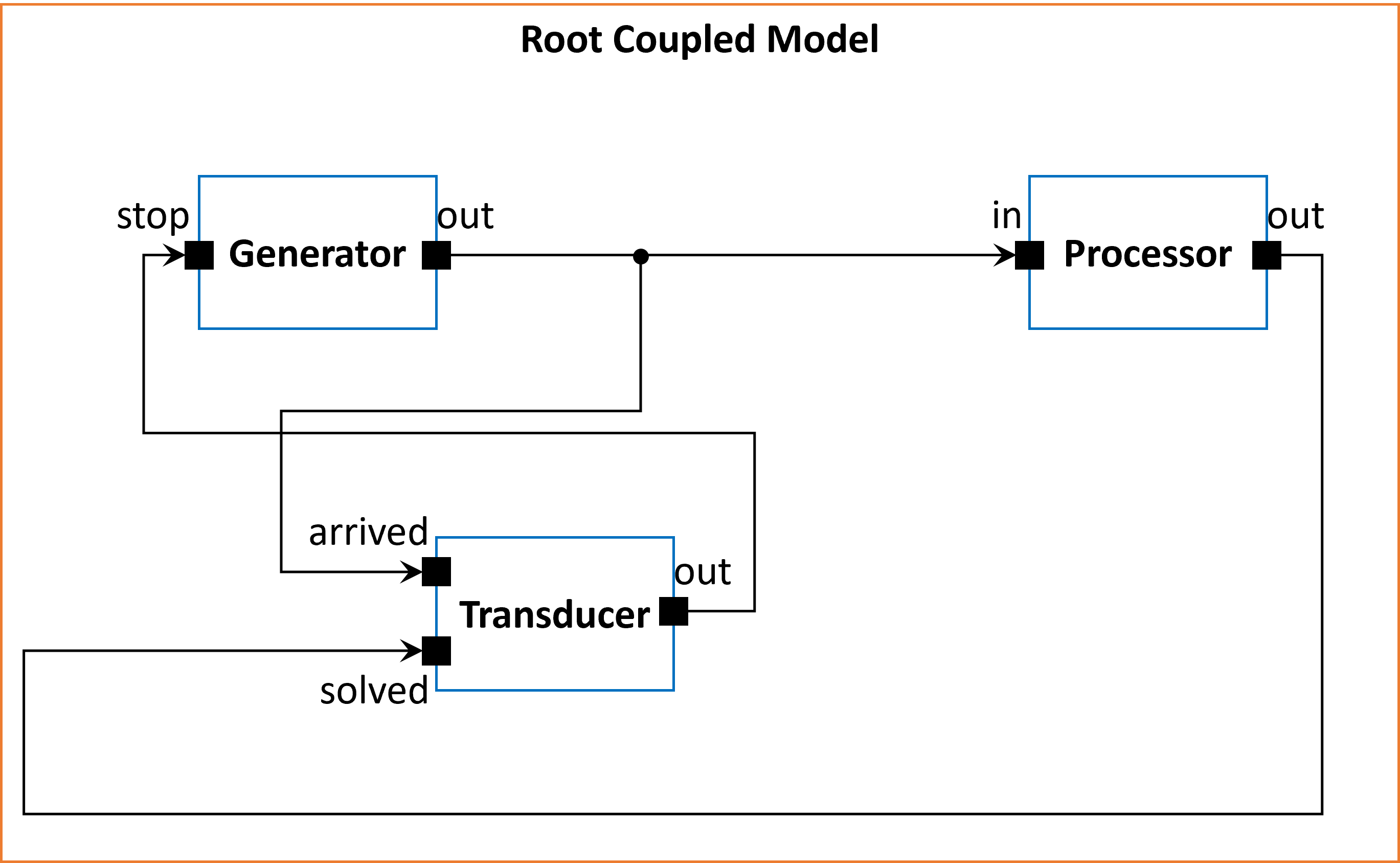}
        \caption{DEVS flattened single-level coupled model}
        \label{fig:gpt}
    \end{subfigure}
    \caption{DEVS coupled and equivalent flattened model.}
    \label{fig:efp_gpt}
\end{figure}

The microservices-based xDEVS distributed simulation execution is explained with the help of the classic Experimental Frame - Processor (EF-P) model \cite{C-Mittal2017}. This model, represented in Fig. \ref{fig:efp},  contains two components: the Experimental Frame (EF) coupled model and the Processor (P) atomic model. As mentioned above, coordinators and simulators are used to specify the structure of a simulation. Each model (or atomic component) is associated with a component simulator. In the case of being a coupled model, it is associated with a component coordinator. In order to simulate it in the Cloud, this hierarchical model is automatically flattened by xDEVS\footnote{The root model is flattened by default. However, there are mechanisms to distribute a non-flattened model using the \texttt{Coupled2Atomic} wrapper \cite{J-Mittal2009a}.}, removing all the intermediate coupled models, in order to obtain the single level coupled model comprising of 3 atomic models: Generator - Processor - Transducer (GPT). The equivalent model depicted in Fig. \ref{fig:gpt}. 

Using a configuration file\footnote{The configuration file enumerates the atomic models and the IP and port where each model is listening, with an equivalent structure to the parallel configuration file}, the distributed simulation can be started by typing in the simulation entities\footnote{With entity we refer to a computer, virtual machine, container, etc. any virtual or physical device able to simulate an xDEVS model} anything equivalent to the following calls (Listing \ref{lst:xDevsDistributed}):

\begin{lstlisting}[language=bash,label={lst:xDevsDistributed}, caption={Execution of a distributed simulation},captionpos=b]
$ # Simulation Entity 2, generator
$ java -cp xdevs.core.simulation.SimulatorDistributed gpt.xml generator
$ # Simulation Entity 3, processor
$ java -cp xdevs.core.simulation.SimulatorDistributed gpt.xml processor
$ # Simulation Entity 4, transducer
$ java -cp xdevs.core.simulation.SimulatorDistributed gpt.xml transducer
$ # Simulation Entity 1, run the Coordinator
$ java -cp xdevs.core.simulation.CoordinatorDistributed gpt.xml
\end{lstlisting}

\begin{figure}
\centering
\includegraphics[width=0.80\textwidth]{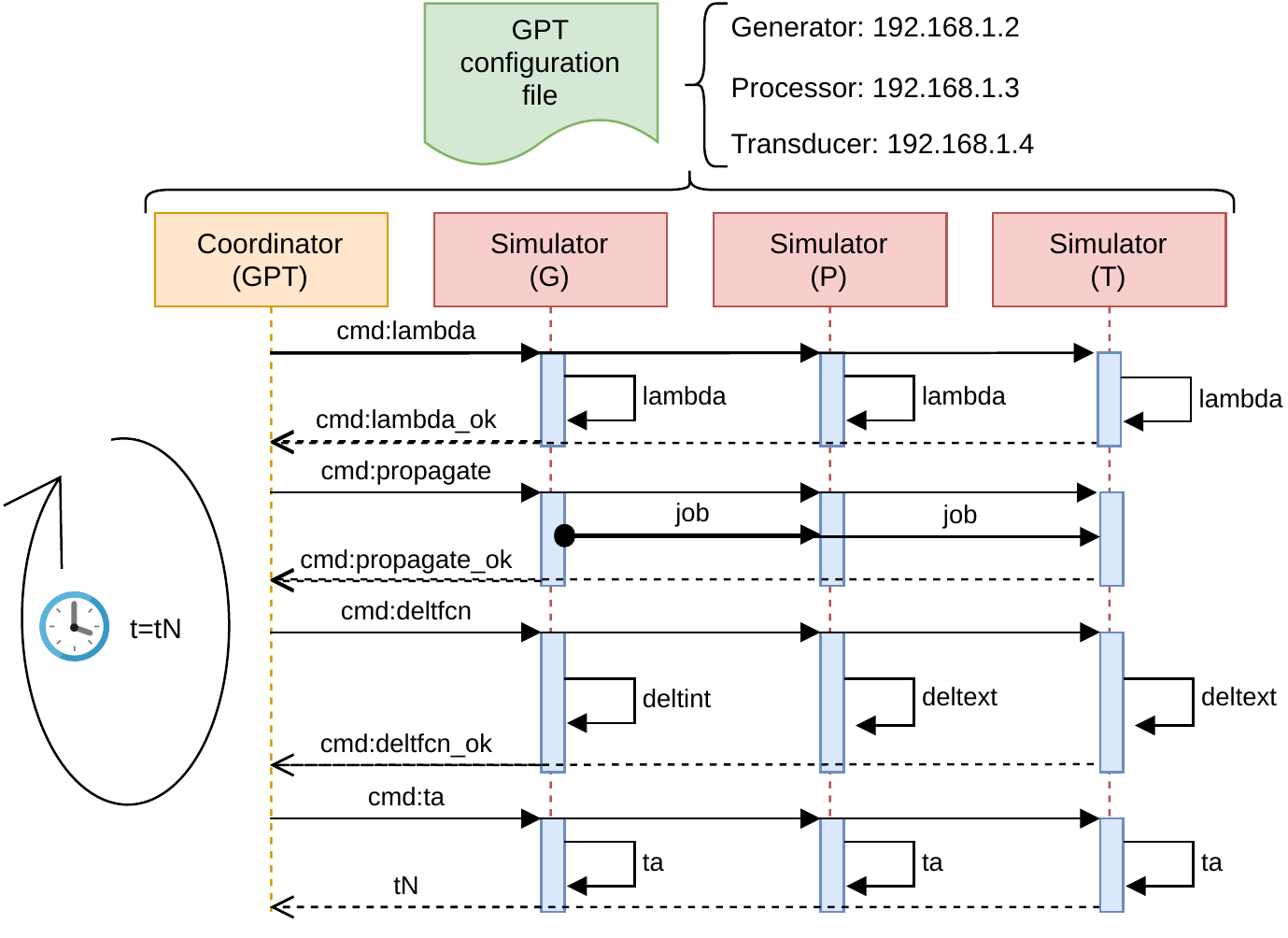}
\caption{Sequence diagram of the xDEVS distributed simulation based on DEVS abstract simulation protocol.}
\label{img:seq_diagram}
\end{figure}

Agnostic of the cloud deployment, a distributed simulation can be seen as a set of independent processes interconnected through the execution of microservices (wrapping DEVS atomic models) that are requested through socket commands. Figure \ref{img:seq_diagram} illustrates the process.

Once the coordinator has been launched, it invokes a command via sockets that executes the output function as a microservice. Each component simulator listens to this command and runs the output function (lambda) of their respective atomic models\footnote{$\lambda$ is executed if and only if the simulation clock is equal to the next time event, according to the DEVS formalism}.  Second, the coordinator invokes the command for the propagation of the output, sent and executed by all the component simulators. To avoid further overheads derived from the network communication, value propagation is performed directly between component simulators without the coordinator acting as a relay between them. After the output propagation, the execution of the transition function is requested, and each component simulator evaluates if the transition function must be the external, internal, or confluent function, depending on the current simulation time, the state and the external input message at the input ports. Finally, the next time event ($tN$ in Figure \ref{img:seq_diagram}) is requested to start the DEVS simulation loop again. This is executed until the number of DEVS iterations is reached, or all the models enter into a \textit{passive} state (i.e., $\sigma = \infty$).

As can be seen, the distributed simulation algorithm is based on the fundamentel DEVS abstract simulation protocol provided in \cite{Zeigler2000}. The model is always the same in the sequential, parallel and distributed execution, and consequently, the current xDEVS architecture unifies the parallel and distribution simulation of Parallel DEVS formalism within the xDEVS implementation. 

\subsubsection{Software architecture}

\begin{figure}[ht]
\centering
\includegraphics[width=0.90\textwidth]{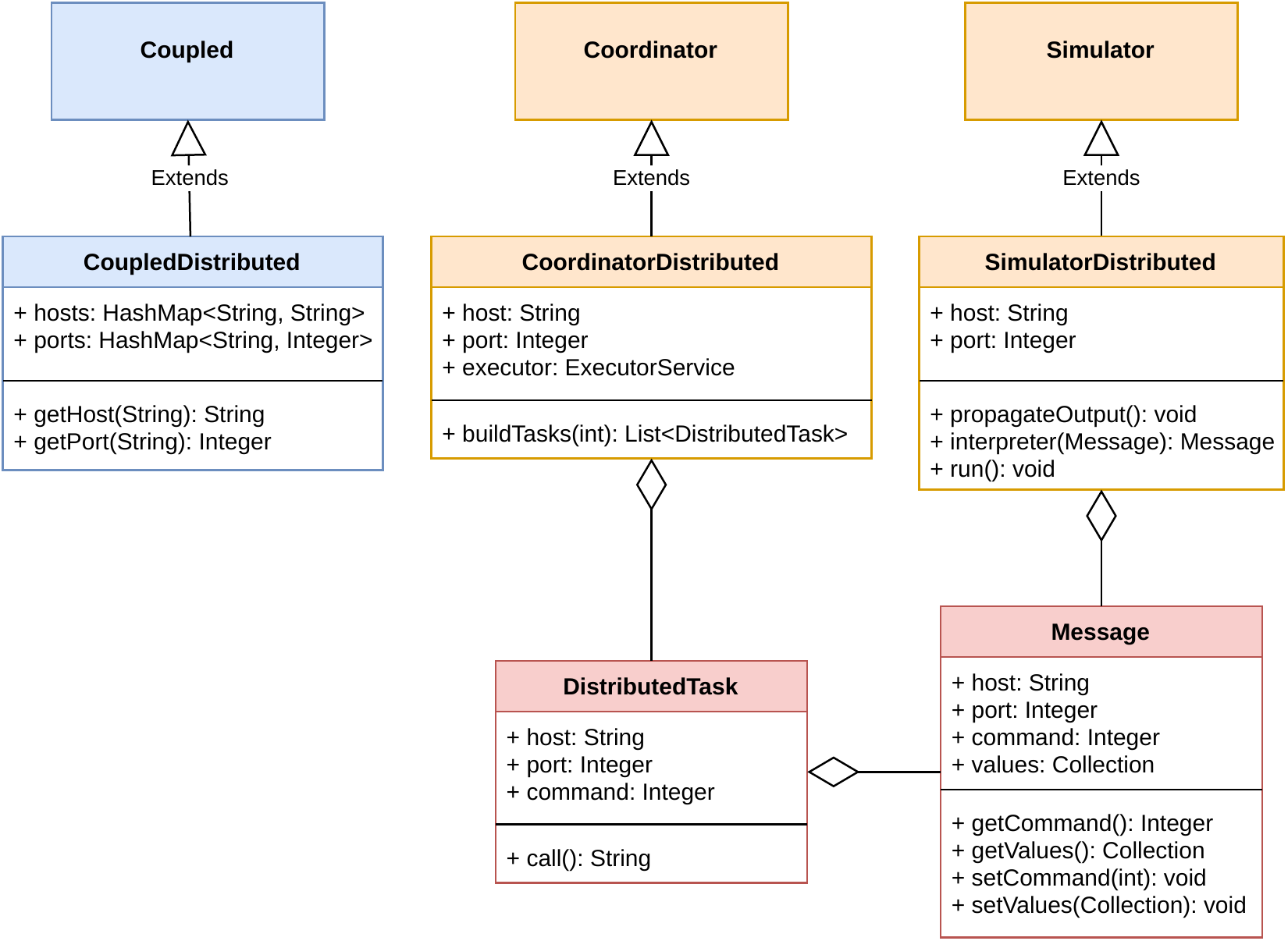}
\caption{Class diagram to support distributed simulation in xDEVS.}\label{fig:distributed_class_diagram}
\end{figure}

The design (Figure \ref{fig:distributed_class_diagram}) is based on a traditional distributed architecture in which each client/server is able to listen, answer and process messages independently and concurrently. The distributed implementation follows the DEVS specification. The coupled model is represented through the \texttt{CoupledDistributed} class, which is the coupled model but with host and port labels into each component. Simulator and coordinator are implemented with the \texttt{CoordinatorDistributed} and \texttt{SimulatorDistributed} classes, respectively. The \texttt{Message} class is implemented to handle the commands sent between coordinator and simulators (see Figure \ref{img:seq_diagram}) and the content is propagated through the ports (via sockets). Finally, the \texttt{DistributedTask} class has been designed to perform all the coordinator tasks in parallel.

The distributed simulation engine does not need additional libraries or frameworks and its deployment can be easily automated, as described in the next section.

\section{Deployment}\label{sec:deployment}
\begin{figure}
    \centering
    \begin{subfigure}{0.45\textwidth}
        \includegraphics[width=\textwidth]{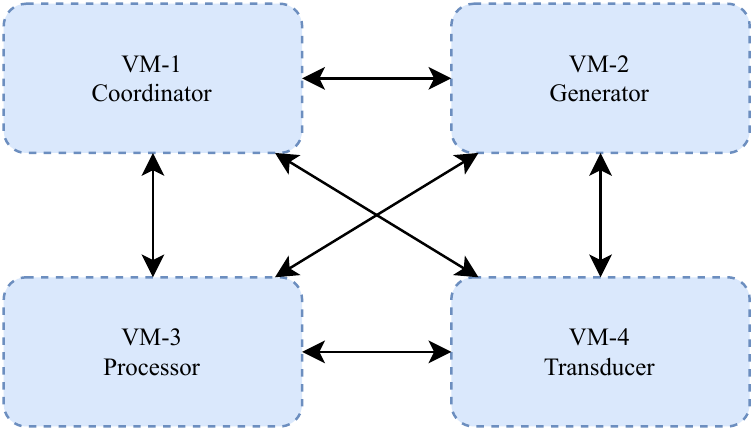}
        \caption{Virtual Machines}
        \label{fig:vms_only}
    \end{subfigure}\\
    \begin{subfigure}{0.45\textwidth}
        \includegraphics[width=\textwidth]{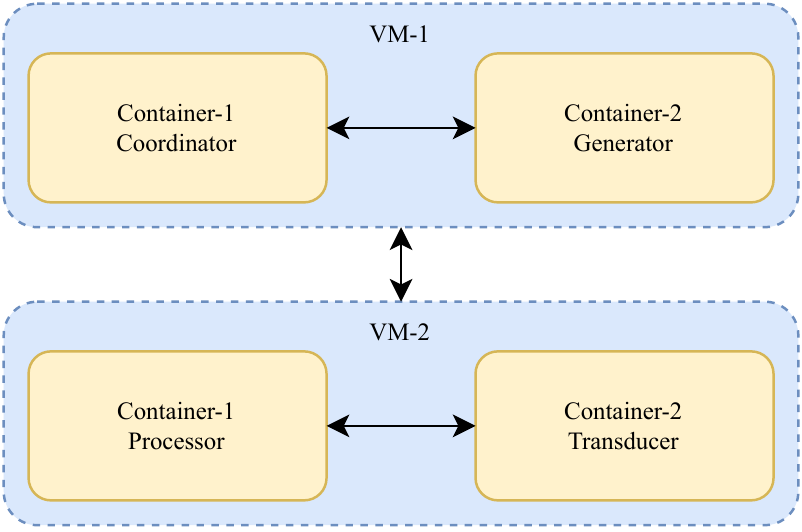}
        \caption{Virtual Machines and Containers}
        \label{fig:vms_containers}
    \end{subfigure}\hfill
    \begin{subfigure}{0.45\textwidth}
        \includegraphics[width=\textwidth]{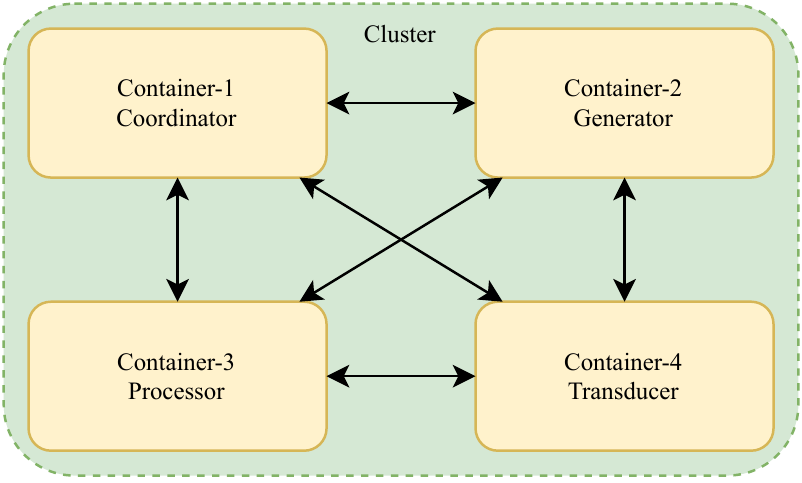}
        \caption{Kubernetes Cluster}
        \label{fig:cluster_kubernetes}
    \end{subfigure}
    \caption{Possible  architectures  for  the  deployment  of  distributed  simulations.}
    \label{fig:distributed_architectures}
\end{figure}

Both the parallel and distributed simulation can be deployed in any computational environment with shared memory. In the case of distributed simulations, each atomic model is executed inside its corresponding component simulator as an isolated process, while a coordinator process marks the beginning and end of the simulation as described in Figure \ref{img:seq_diagram}. The communication between these simulators is performed using network sockets. As a result, any distributed architecture is possible. Figure \ref{fig:distributed_architectures} shows some examples. For instance, Figure \ref{fig:vms_only} illustrates a simple GPT deployment using only Virtual Machines, a more traditional approach. Figure \ref{fig:vms_containers}, on the other hand, illustrate the same distribution but with containers inside the Virtual Machines. Finally, the example shown in Figure  \ref{fig:cluster_kubernetes} is the one used in this work, where the set of containers are managed by a kubernetes cluster. These architectures are possible and feasible to deploy in the cloud using services provided by infrastructure providers that are well known to date: Google, Amazon and Microsoft Azure among others, and whose services are similar or at least use standard virtualization tools such as Docker and Kubernetes. For the purposes of this research, we have selected the Google Cloud Platform services, in particular we have used for the parallel simulations a single virtual machine, and for the distributed simulations a cluster of containers automatically deployed through the Google Kubernetes Engine (GKE).

\begin{figure}
\centering
\includegraphics[width=0.90\textwidth]{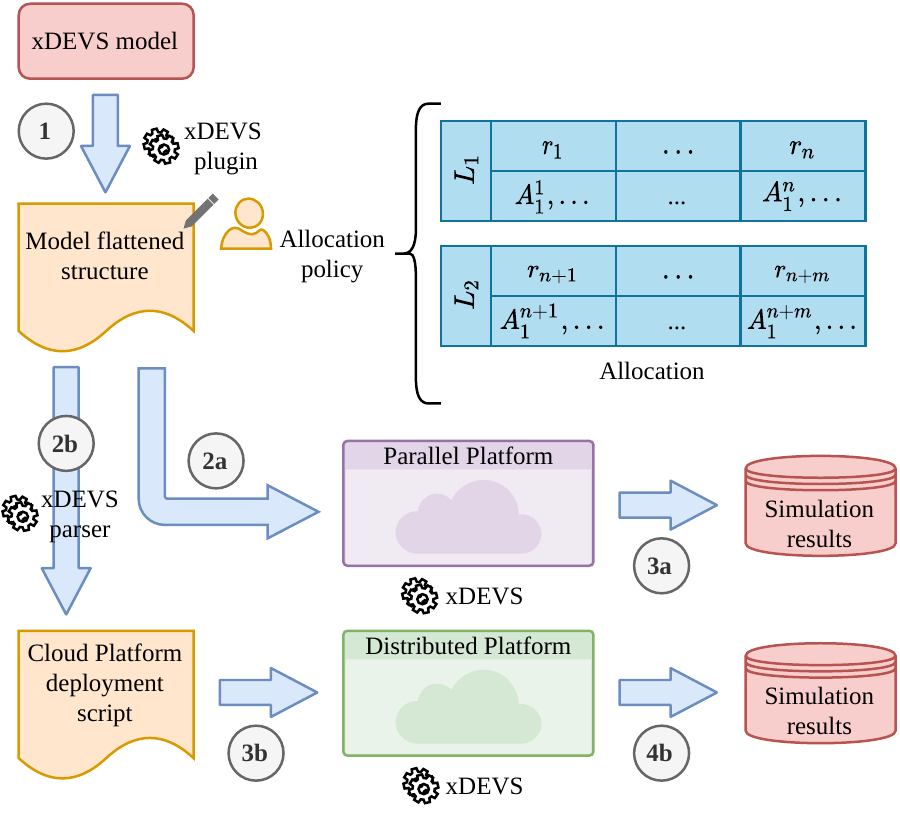}
\caption{Cloud deployment scheme}
\label{fig:cloud_deployment}
\end{figure}

Figure \ref{fig:cloud_deployment} shows the steps that must be followed to execute a parallel or distributed simulation. This process is derived from our earlier DEVS/SOA deployment mechanisms \cite{J-Mittal2009a}.

In the first phase, an XML description of a flattened version of the original model is generated with xDEVS. This text file contains all the atomic models and coupling relations obtained after rearranging the connections of the coupled models, which are removed by default to facilitate the deployment~\cite{risco2008optimization} and reduce simulation overheads. This text file, in addition to the traditional DEVS attributes (component's names, port names, connections, etc.), also contains a host address that identifies a simulation entity (to be deployed in a computational execution environment), and a communication endpoint (named port as well) for the case of distributed simulation deployment, and a thread pool name for the parallel simulation deployment. Since the generation of the text file is automated, it generates a single host name and endpoint or a single thread pool. However, this file can be edited to change the default behavior. 

Although all the atomic models can be allocated to a single container or a thread pool, this option is not yet operational since a DEVS model can contain hundreds of atomic models, with a huge variety of computational weight in terms of CPU cycles. Thus, editing this initial text file, as Figure \ref{fig:cloud_deployment} shows, allows us to group several atomic models per container set (distributed deployment) or thread pool (parallel deployment). Figure~\ref{fig:cloud_deployment} shows a 2-level allocation policy used in this paper. This allocation distributes the atomic models ($A_i^j \in \mathbf{A}$) over two container sets or two thread pools. Those with high computational demands are placed at level 1 ($L_1$), from $r_1$ to $r_n$. The remaining atomic models are distributed over the level 2  ($L_2$), from $r_{n+1}$ to $r_{n+m}$. Here $r_i$ represents a computational resource. $r_i$ is a container in the case of the distributed simulation following the scheme provided in Figure \ref{fig:cluster_kubernetes}, or a single thread in the case of a parallel simulation. In any case, one or more atomic models can be allocated in each resource. In general, $n > m$ in order to follow a coarse-grain allocation policy that:
\begin{enumerate}
    \item exploits the modeler knowledge about which atomic models consume more CPU, and
    \item avoids a computing-intensive profiling phase.
\end{enumerate}

The second phase, after the allocation policy is completed, depends on the simulation type. In the parallel case, the model is just simulated with the \texttt{CoordinatorParallel} class (see step \texttt{2a} in Figure \ref{fig:cloud_deployment}), creating the specified thread pools, and the simulation results are obtained. In the distributed case, a parser reads the XML file and generates an architecture-specific script as a YAML deployment file (see step \texttt{2b} in Figure \ref{fig:cloud_deployment}). This file describes the distributed simulation deployment structure, including the pods configurations, their inner containers, and the ports opened in these containers to communicate the different models over the network. In this case, we specify  single-container pods. Note that the design of this parser is straightforward, simply consists on reading an XML file and generating a YAML file, and can be adapted to other service providers.

In the third (distributed) phase, the pods specified in the YAML file are created in the selected cloud platform, and the model is deployed as described by the allocation policy. In this step, the atomic models are distributed over the containers, instantiating the suitable simulator processes per the DEVS simulation protocol. Therefore, each container executes one or more distributed xDEVS simulators, each one with its corresponding atomic model. Besides, one particular container runs the distributed xDEVS root coordinator. Typically, once the simulation ends, the results are stored in a distributed way, as each atomic model can have a different mechanism to save its data. A recommended approach for unifying these data  is to have different Transducer atomic models~\cite{Zeigler2000}, collecting the relevant information and storing it in the suitable repositories.

The main difference between the parallel and distributed simulations is that in the case of parallel simulations, two or more thread pools are executed sequentially, one after the other, although each thread pool is parallel of course. In the case of the distributed simulation, each simulator is an independent full process, which demands a lot of dedicated memory, but the simulation is intrinsically parallel, independent of the number of containers used.

\section{Evaluation}\label{sec:evaluation}
In this section, we evaluate both the parallel and distributed coordinators of the xDEVS simulation engine. This is performed through the DEVStone benchmark. To the best of our knowledge, this is the first time the DEVStone benchmark is used with a delay in the transition functions to measure the performance of discrete event simulation engines. Including the delay aspect is essential to evaluate the impact of model's execution on CPU load. We first describe the DEVStone benchmark and how the delay is introduced. Next, we perform an analysis of synthetic delay distribution selecting a DEVStone model class to assign different delay weights to the set of atomic models. Once the delays are assigned, we proceed with the analysis of the parallel and distributed simulations, and provide the comparison results.

\subsection{The DEVStone benchmark}

DEVStone \cite{Glinsky2005} is a synthetic benchmark devoted to automating the evaluation of DEVS-based simulation approaches. It allows the generation of different types of models, each of them specialized in measuring specific aspects of the simulation. This benchmark has become popular over the years, and has been used extensively in literature to evaluate and compare the performance of different DEVS simulators \cite{J-RiscoMartin2017, VanTendeloo2014}. 

DEVStone describes several synthetic models that can be configured to vary their size and complexity. With this aim, a recursive structure with configurable depth where all the levels contain equivalent components and interconnections is presented. The customization of the models is done through the use of four parameters: (i) \textit{width}, that affects to the number of components (1 coupled and $width-1$ atomic models) per layer, (ii) \textit{depth}, that specifies the number of nested coupled models, (iii) \textit{internal transition delay}, and (iv) \textit{external transition delay}. According to the DEVStone specifications, these two delay times are spent executing Dhrystones~\cite{Weicker1984} to keep the CPU busy. It is worthwhile to mention that in this work we compute this delay as \textbf{CPU time}. i.e., the Dhrystone benchmark loop is executing iterations as long as the CPU time consumed (not the wall clock time) is less than $\Delta_\mathrm{int}$ in the internal transition function, or $\Delta_\mathrm{ext}$ in the external transition function. It is important to measure CPU time because otherwise the CPU can run hundreds of simultaneous transition functions consuming the corresponding $\Delta_\mathrm{int}$ and $\Delta_\mathrm{ext}$ wall clock delays, and not being forced to keep each transition function in the CPU for the specified time.

The behavior of a DEVStone model is conducted by the distribution of its DEVStone atomic models. The DEVS specification of a DEVStone atomic model is shown in Algorithm~\ref{alg:atomic}. 

\begin{algorithm}
\caption{DEVStone atomic model
\label{alg:atomic}}
\begin{algorithmic}

\REQUIRE{\texttt{NUM\_DELT\_INTS}, \texttt{NUM\_DELT\_EXTS} and \texttt{NUM\_OF\_EVENTS} are global variables, and store the total number of internal transition functions, external transition functions and events triggered inside the whole model. $\Delta_{\rm int}$  and $\Delta_{\rm ext}$ are the delays introduced in the internal and external transition functions, respectively.}

\textbf{function} [list,\emph{phase},$\sigma$] = init()
\STATE list = [] \COMMENT{list is part of the state, and stores all the events received by this atomic model}
\STATE $\sigma = \infty$

\textbf{function} [list,\emph{phase},$\sigma$] = $\delta_{\rm int}$(list,\emph{phase},$\sigma$)
\STATE \texttt{NUM\_DELT\_INTS} = \texttt{NUM\_DELT\_INTS} + 1
\STATE Dhrystone($\Delta_{\rm int}$)
\STATE list = []
\STATE $\sigma = \infty$

\textbf{function} [list,\emph{phase},$\sigma$] = $\delta_{\rm ext}$(list,\emph{phase},$\sigma$,$e$,$X^b$)
\STATE \texttt{NUM\_DELT\_EXTS} = \texttt{NUM\_DELT\_EXTS} + 1
\STATE Dhrystone($\Delta_{\rm ext}$)
\STATE values = $X^b(in)$ \COMMENT{$X^b(in)$ is a list containing all the events waiting in the ``in'' input port}
\STATE \texttt{NUM\_OF\_EVENTS} = \texttt{NUM\_OF\_EVENTS} + values.size()
\STATE list = [list;values] \COMMENT{Concatenate both lists}
\STATE \emph{phase = ``active''}
\STATE $\sigma = 0$

\textbf{function} [list,\emph{phase},$\sigma$] = $\delta_{\rm con}$(list,\emph{phase},$\sigma$,$ta(s)$,$X^b$)
\STATE $\delta_{\rm ext}$($\delta_{\rm int}$(list,\emph{phase},$\sigma$),0,$X^b$)

\textbf{function} $\lambda ()$
 \STATE send(``out'', list) \COMMENT{sends the whole list by the ``out'' output port}

\textbf{function} $\sigma$ = $ta$(list,\emph{phase},$\sigma$)
 \STATE $\sigma = \sigma$

\end{algorithmic}
\end{algorithm}

\begin{figure}
    \centering
    \begin{subfigure}[b]{0.45\textwidth}
        \includegraphics[width=\textwidth]{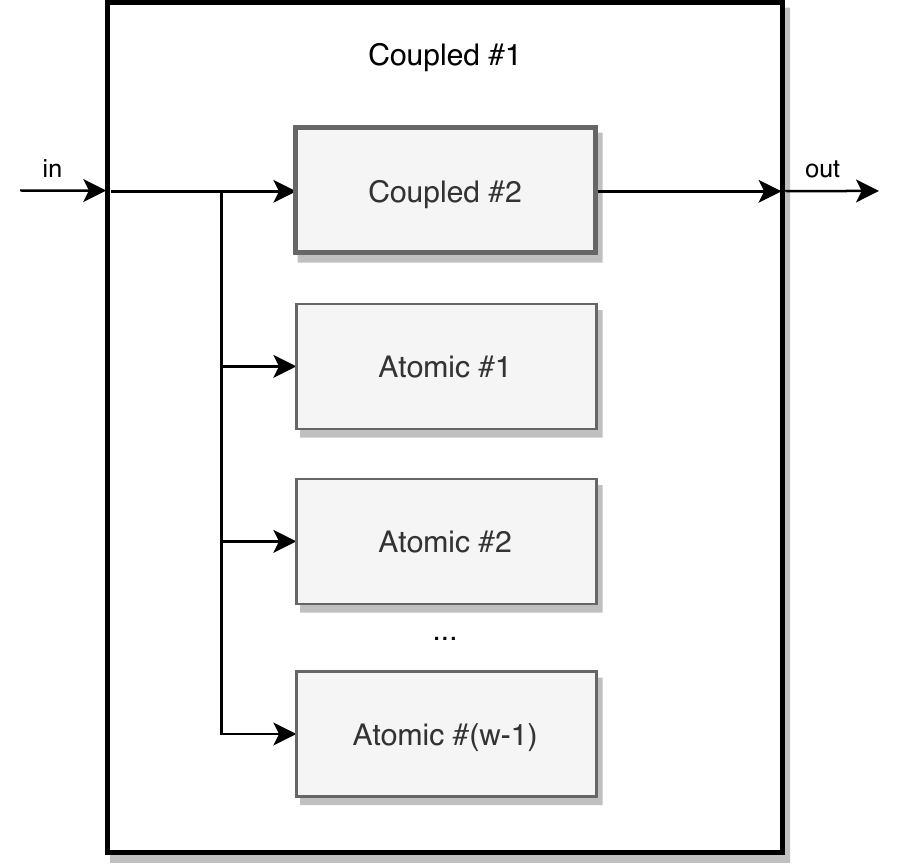}
        \caption{Low level of Interconnections model (LI).}
        \label{img:devstone_li}
    \end{subfigure}\hfill
    \begin{subfigure}[b]{0.45\textwidth}
        \includegraphics[width=\textwidth]{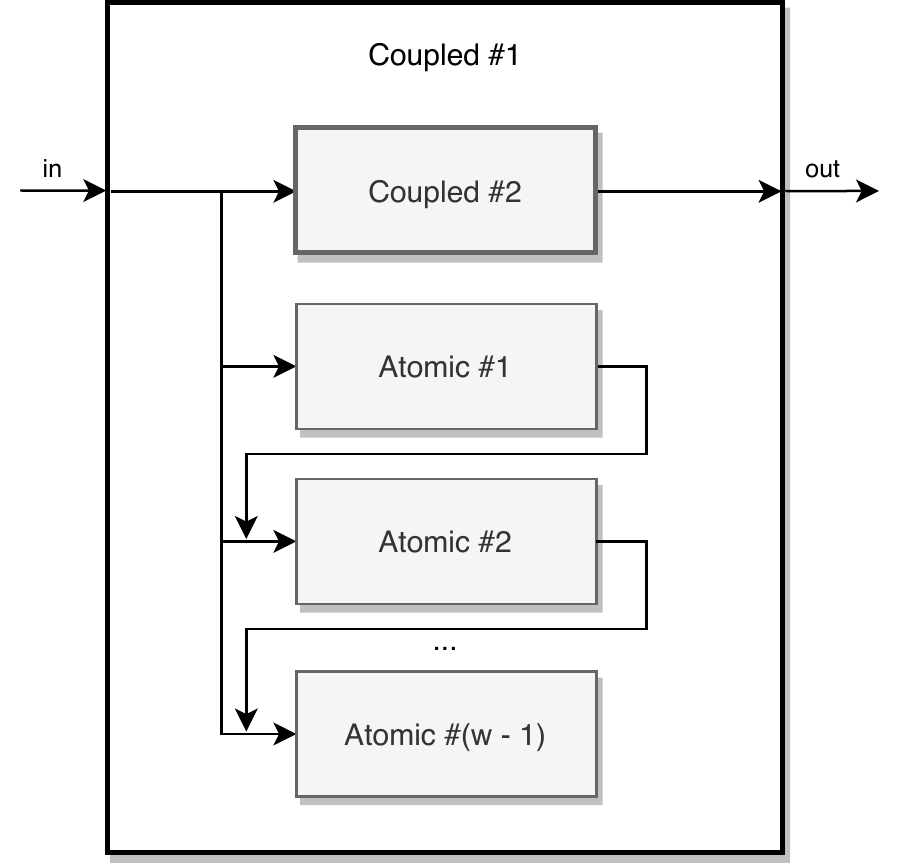}
        \caption{High Input couplings model (HI).}
        \label{img:devstone_hi}
    \end{subfigure}
    \begin{subfigure}[b]{0.46\textwidth}
        \includegraphics[width=\textwidth]{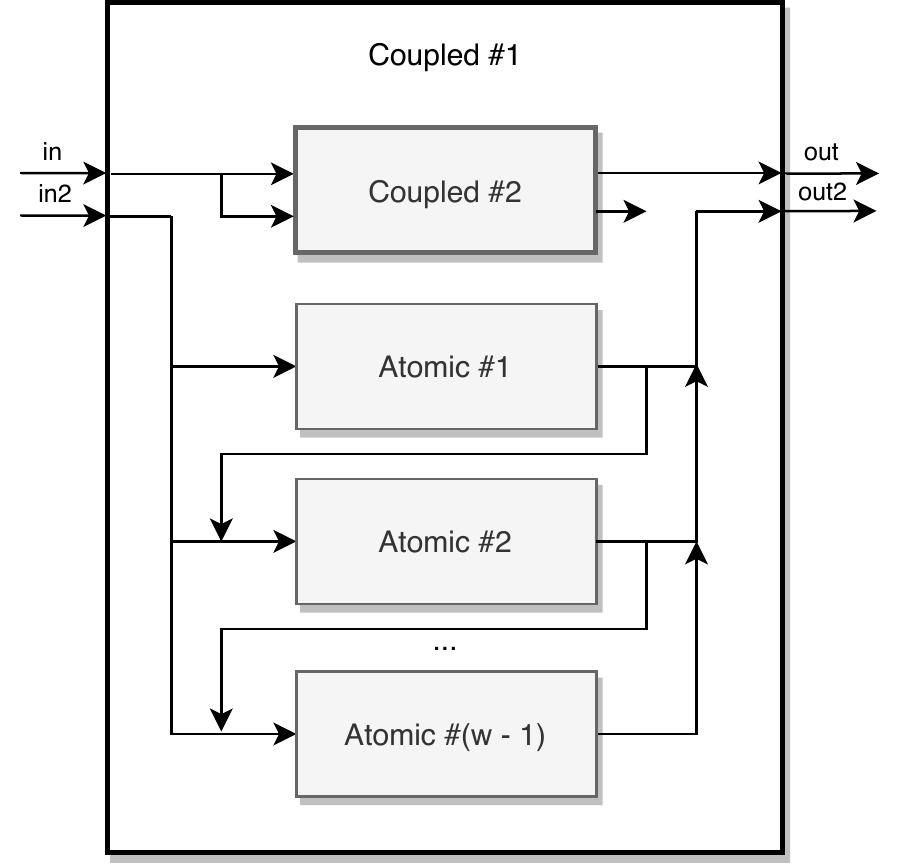}
        \caption{HI model with numerous Outputs model (HO).}
        \label{img:devstone_ho}
    \end{subfigure}\hfill
    \begin{subfigure}[b]{0.42\textwidth}
        \includegraphics[width=\textwidth]{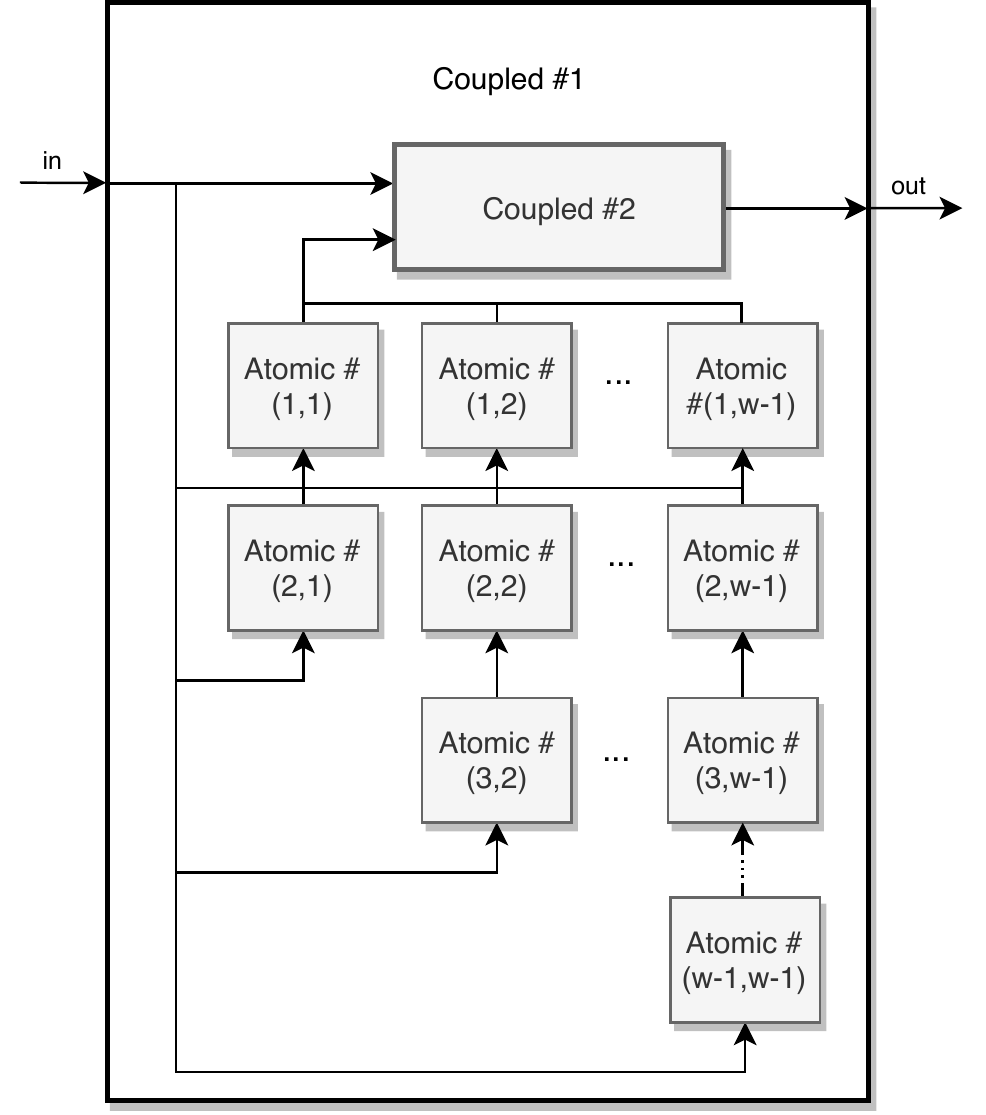}
        \caption{Exponential level of coupling and outputs model (HOmod).}
        \label{img:devstone_homod}
    \end{subfigure}

    \caption[DEVStone models]{DEVStone models internal structure.}
    \label{img:devstone_models}
\end{figure}

DEVStone describes four types of models (depicted in Figure~\ref{img:devstone_models}):

\begin{itemize}
    \item \textbf{LI} (Low level of Interconnections) models are the simplest models, with a low level of coupling relations in their coupled models (Figure~\ref{img:devstone_li}).
    \item \textbf{HI} (High Input couplings) models are similar to LI models, but increases the number of internal couplings (Figure~\ref{img:devstone_hi}).
    \item \textbf{HO} (Hi model with numerous Outputs) models are a variation of the HI models where all the atomic components in each coupled module are connected to the coupled output port. It is worth noting that these models present unconnected ports that may serve to detect malfunctioning in the simulators when cleaning the values of ports without couplings (Figure~\ref{img:devstone_ho}).
    \item \textbf{HOmod} models reproduce an exponential level of coupling and outputs model (Figure~\ref{img:devstone_homod}).
\end{itemize}

Analyzing the publications that study the performance of DEVS simulation engines through DEVStone, we may find that the HO set offers a good balance between CPU and memory usage \cite{J-RiscoMartin2017, Cardenas2020}. As a result, we use the HO set of DEVStone models to evaluate the performance of our DEVS parallel and distributed simulation engines. In HO, the deepest coupled model is formed by one single atomic model. As Figure \ref{img:devstone_ho} shows, the remaining coupled models are constituted by 1 coupled model, a chain of $w-1$ atomic models, and a set of $k = 1 \ldots w-1$ chains formed by $\sum_{i=1}^{k}i$ atomic models. The second external input port is connected to the whole first row and only to the first atomic component in the remaining rows. Additionally, all the atomic models in the second row are connected to the first row, which in turn send the whole output directly to the coupled component. Finally, each remaining atomic component is connected to its upper component. The computation of the total number of atomic models, couplings, executions of transition functions and number of events propagated is quite straightforward \cite{J-RiscoMartin2017}:

\begin{eqnarray}
\#{\rm Atomic} & = & 1 + (d-1) \cdot (w-1) \label{eq:atomics} \\
\#{\rm EIC} & = & 1 + (d-1) \cdot (w + 1) \\
\#{\rm IC} & = & (d-1) \cdot (w-2) \\
\#{\rm EOC} & = & 1 + (d - 1) \cdot w \\
\#\delta_{\rm int} & = & 1+ (d-1) \cdot \sum_{i=1}^{w-1}i \\
    & = & 1 + (d-1) \cdot \frac{w^2 -w}{2} \\
\#\delta_{\rm ext} & = & 1 + (d-1) \cdot \frac{w^2 -w}{2} \\
\#{\rm Events} & = & 1 + (d-1) \cdot \frac{w^2 -w}{2}
\end{eqnarray}

\subsection{Profiling of the benchmarks}

First, HO model size and transition delays must be fixed to reach a good trade-off between the number of atomic models and the simulation time.

To find these values, we have performed the profiling of different HO model parameters. This Section shows the results obtained and proceeds with the selection of one HO model. All the profiling was performed on a virtual machine with 4 Intel(R) Xeon(R) CPU @ 2.8 GHz and 32 GiB RAM, based on the N2 Google Cloud configuration series and the Debian GNU/Linux 10 operating system and with OpenJDK 11. This is the minimum node able to run distributed simulations, so it is fixed as the base machine for sequential, parallel and distributed experiments.

For the sake of clarity we have firstly considered \emph{squared} models, i.e., HO width equal to HO depth ($w=d$). We have tested six different sizes $w=10,11,\ldots , 15$, which leads to 82, 101, 122, 145, 170, and 197 atomic models respectively, following (\ref{eq:atomics}). Next, we have defined the external transition delay equal to the internal transition delay in each atomic model ($\Delta_\mathrm{int}^{i \in \mathbf{A}}=\Delta_\mathrm{ext}^{i \in \mathbf{A}}=\Delta_{i \in \mathbf{A}}$). Then, each $\Delta_i$ has been defined using the following seven configurations: 

\begin{itemize}
    \item A constant value for all the atomic models: $\Delta_{i \in \mathbf{A}} = k$ seconds. We have performed three tests with $k=1,2,3$.
    \item A random value using a uniform real distribution: $\Delta_{i \in \mathbf{A}} = N(0,k)$ seconds. Again, we have performed three tests with $k=1,2,3$
    \item A random value using a chi square distribution: $\Delta_{i \in \mathbf{A}} = \chi^2(f)$ seconds, with $f=2$.
\end{itemize}

\begin{figure}
\centering
\includegraphics[width=0.85\textwidth]{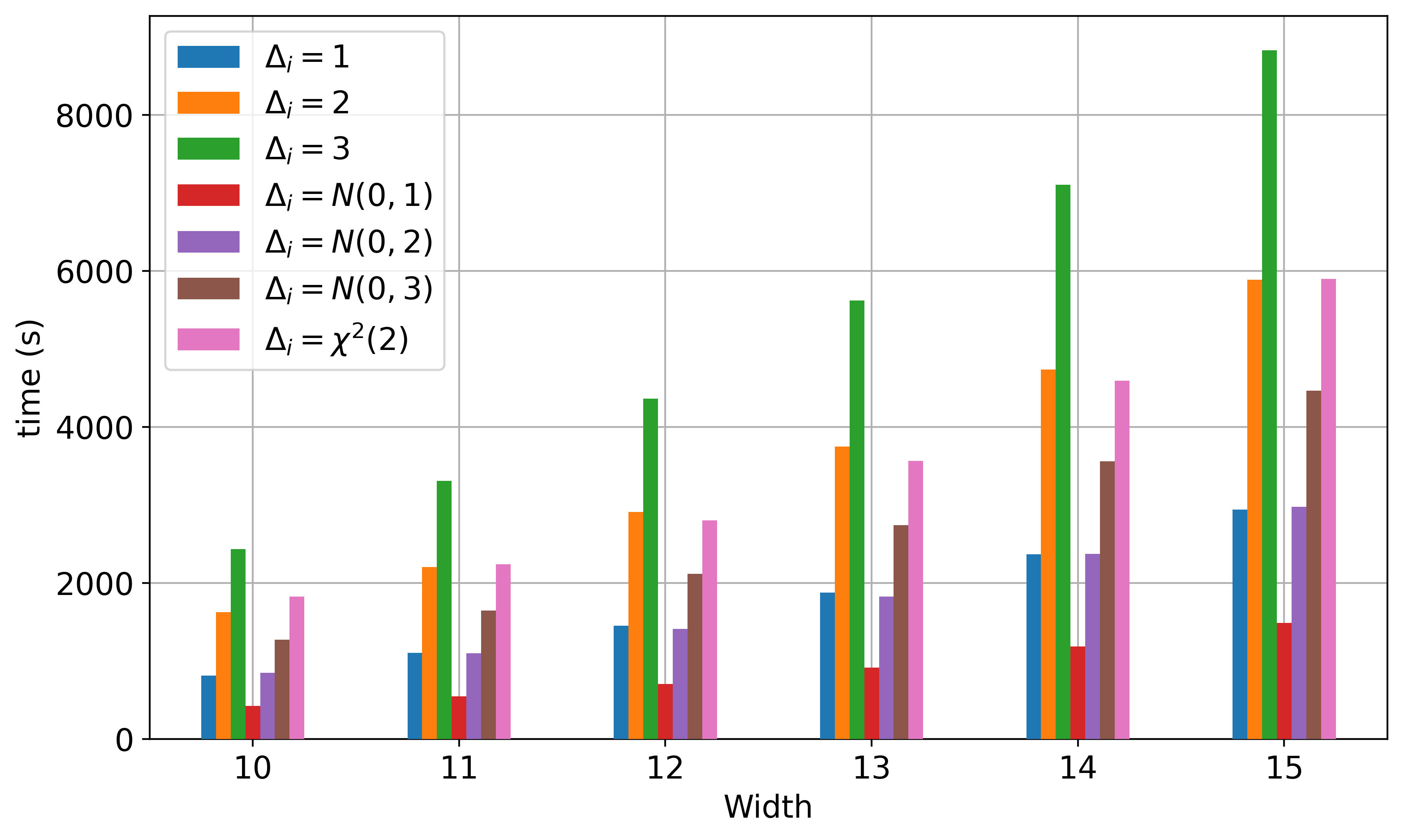}
\caption{HO simulation times.}
\label{fig:sim_times}
\end{figure}

Figure \ref{fig:sim_times} depicts the simulation times of the six different HO model sizes for the seven different configuration of the transition delays. In order to allow the repetition of parameters configuration in the parallel and distributed experiments, we have fixed the random seed. With independence of the HO model size, slowest simulations corresponded to $\Delta_i = 3$, followed by $\Delta_i = \chi^2(2)$ or $\Delta_i = 2$, and $\Delta_i = N(0,3), \Delta_i = N(0,2), \Delta_i = 1$, and $\Delta_i = N(0,1)$.

HO models with $\Delta_i = 3$ and $w = 15$ were simulated in $8826.06$ seconds. On the other hand, HO models with $\Delta_i = N(0,1)$ and $w=10$ were simulated in $424.58$ seconds. As a result, HO models with width equal to $15$ seem to be an appropriate model size (197 atomic models) to perform different parallel and distributed simulations with a significant number of thread pools and containers, respectively. To select one of the seven distributions for the delays, we have checked the simulation times consumed by each atomic model. To this end, we have examined $\Delta_i=2, \Delta_i=N(0,3)$, and $\Delta_i=\chi^2(2)$, since they offer equivalent simulation times (range 4000-6000 seconds) and represent the three distribution classes (constant, uniform and chi square). 

\begin{figure}
    \centering
    \begin{subfigure}{0.45\textwidth}
        \includegraphics[width=\textwidth]{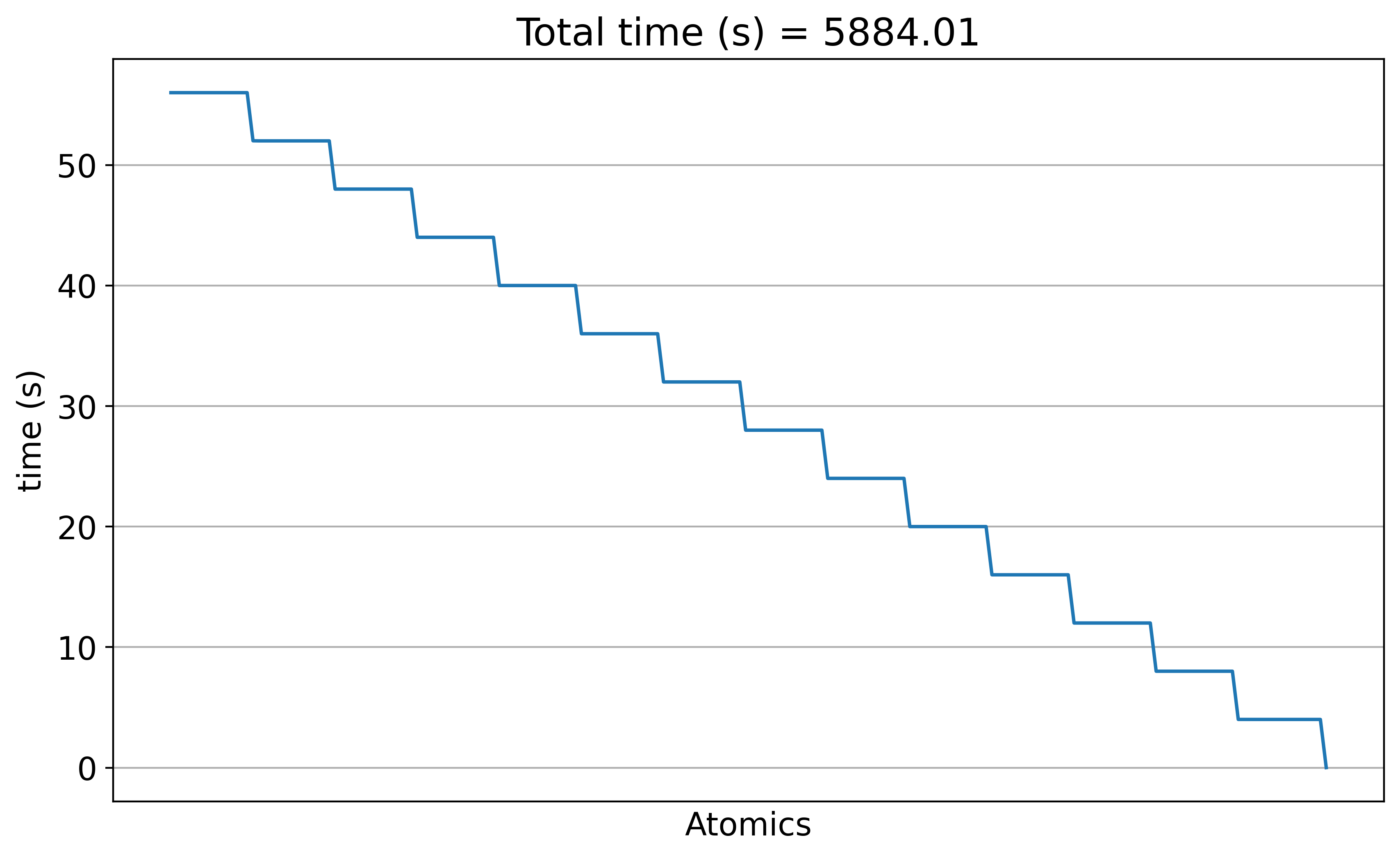}
        \caption{$\Delta_i=2$}
        \label{fig:delta-cte-2}
    \end{subfigure}\hfill
    \begin{subfigure}{0.45\textwidth}
        \includegraphics[width=\textwidth]{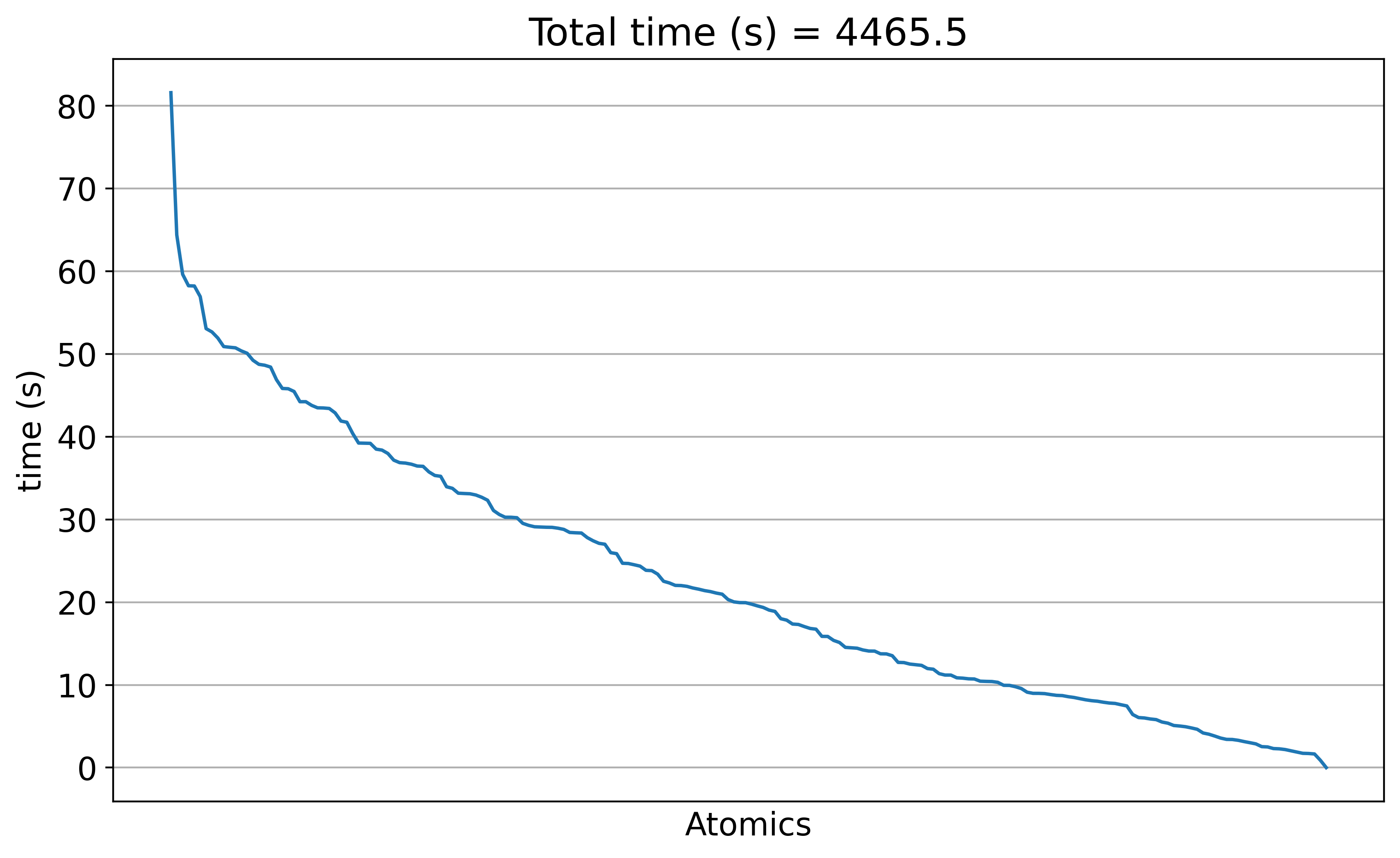}
        \caption{$\Delta_i=N(0,3)$}
        \label{fig:delta-uni-3}
    \end{subfigure}
    \begin{subfigure}{0.45\textwidth}
        \includegraphics[width=\textwidth]{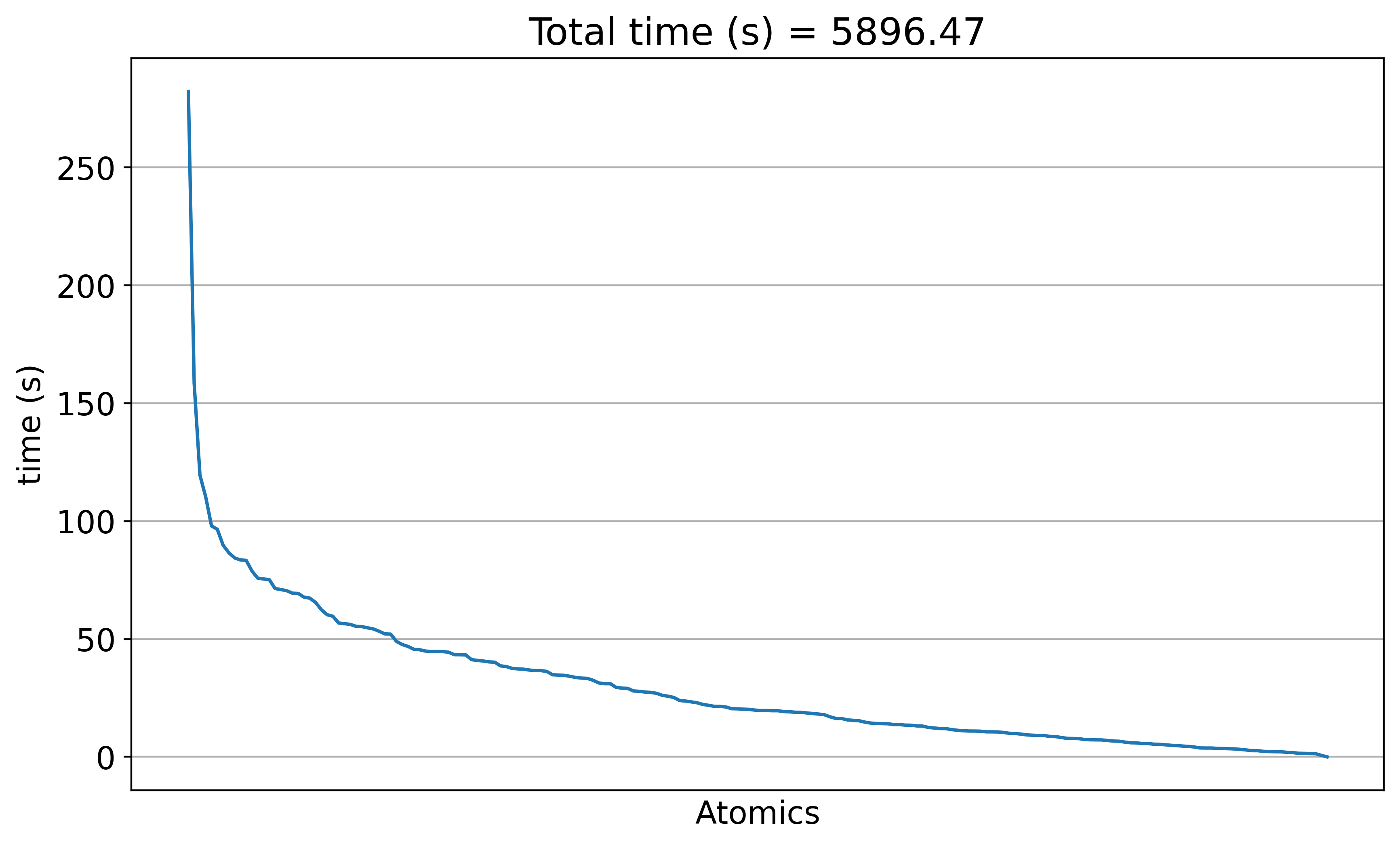}
        \caption{$\Delta_i=\chi^2(2)$}
        \label{fig:delta-chi-2}
    \end{subfigure}
    \caption{Simulation times consumed by the atomic models using three different distributions and constant size ($w=15$).}
    \label{fig:distribution-profiles}
\end{figure}

Figure \ref{fig:distribution-profiles} illustrate the results. For the sake of clarity, we have not labeled the atomic models, but have ordered them from higher to lower simulation time consumed by each one. Table \ref{tab:distribution-profiles} shows the most representative atomic models of each distribution. Following the numbering scheme in Figure \ref{img:devstone_models}, an atomic component is labeled as $A_i^j$, being $j$ the coupled model where the atomic model belongs to (with $1$ the root HO coupled model and $d=w=15$ the last one), and $i \in {1 \ldots w-1=14}$ the number of the atomic component in the $j-th$ coupled component's chain of models.

 The constant distribution ($\Delta_i=2$) gives a well-known behavior. Since each atomic model $A_i$ receives $i$ events in cascade, a total of $i$ external and internal transitions functions are executed ($2 \times i $ transitions). Thus, the simulation time consumed by the transitions functions of $A_i$ is approximately $2\times i \times \Delta_i = 4 \times i$. There are then $14$ atomic models $A_{14}^*$ consuming  $56$ seconds, $14$ atomic models $A_{13}^*$ consuming $52$ seconds, and so forth.

\begin{table}
    \centering
    \begin{tabular}{|c|c|c|}
         \textbf{Name} & \textbf{$t_{i \in \mathbf{A}}$ = time($\delta_\mathrm{ext}$)+time($\delta_\mathrm{int}$)} & \textbf{\% of $\sum{t_i}$} \\
         \hline
        \multicolumn{3}{|c|}{$\Delta_i=2$} \\
        \hline
        $A_{14}^{7}$	& 56.00	& 1 \\
        $A_{14}^{1}$	& 56.00	& 1 \\
        $A_{14}^{13}$	& 56.00	& 1 \\
        $A_{14}^{12}$	& 56.00	& 1 \\
        $\ldots$ & $\ldots$ & $\ldots$ \\
        $A_{1}^{10}$    & 4.00	& 0 \\
        $A_{1}^{5}$	    & 4.00	& 0 \\
        $A_{1}^{7}$ 	& 4.00	& 0 \\
         \hline
        \multicolumn{3}{|c|}{$\Delta_i=N(0,3)$} \\
        \hline
        $A_{14}^{5}$    & 81.58 & 2 \\
        $A_{12}^{6}$    & 64.38	& 1 \\
        $A_{14}^{13}$   & 59.63	& 1 \\
        $A_{13}^{8}$    & 58.24	& 1 \\
        $\ldots$ & $\ldots$ & $\ldots$ \\
        $A_{1}^{9}$     & 1.71	& 0 \\
        $A_{1}^{10}$    & 1.65	& 0 \\
        $A_{1}^{14}$    & 0.89	& 0    \\     
         \hline
        \multicolumn{3}{|c|}{$\Delta_i=\chi^2(2)$} \\
        \hline
        $A_{14}^{5}$    & 282.20 & 5 \\
        $A_{14}^{3}$    & 158.42 & 3 \\
        $A_{13}^{8}$    & 119.40 & 2 \\
        $A_{12}^{6}$    & 110.19 & 2 \\
        $\ldots$ & $\ldots$ & $\ldots$ \\
        $A_{1}^{1}$     &	1.42 & 0 \\
        $A_{2}^{5}$     &	1.37 & 0 \\
        $A_{1}^{14}$    &	0.65 & 0 \\        
    \end{tabular}
    \caption{Simulation times consumed by the most representative atomic models using three different distributions and constant size ($w=15$).}
    \label{tab:distribution-profiles}
\end{table}

The uniform distribution $N(0,3)$ gives a minimum value of $0$, a maximum value of $3$, and a mean value of $1.5$, equally distributed. Thus, we can expect a maximum consumption of $2 \times 14 \times 3 = 84$ seconds, close to $t(A_{14}^5)=81.58$ seconds in Table \ref{tab:distribution-profiles}, and then a smooth linear drop, equivalent to the one observed in the constant distribution.

The chi square distribution gives a minimum value of $0$, statistically a maximum value of $10.60$, and a mean value of $2$. There are a few models with high $\Delta_i$ values, since the distribution is slightly unbalanced. We can expect a maximum consumption of $2 \times 14 \times 10.60 = 296.80$ seconds, again close to the $t(A_{14}^6)=282.20$ seconds in Table \ref{tab:distribution-profiles}, and then a brief abrupt drop, followed by a smooth descend.

Since the chi square distribution shows more variability, it covers pretty well the spectrum of simulations we would like to analyze. Having such variety of simulation times, we can better analyze the impact on the number of threads or containers deployed for the distributed simulation, as well as the number of distributed atomic models allocated in them.

\subsection{Parallel simulation}

In the following we show the results obtained by the parallel simulations. To this end, we have configured several experiments following the deployment illustrated in Figure \ref{fig:cloud_deployment} with two thread pools, and additionally a parallel execution with a single thread pool. The hardware resources management for each thread pool has been left to the operating system.

We have used the distributed simulation as a reference to set up the baseline virtual machine. As a consequence, we have tested the parallel simulations using a 4 Intel(R) Xeon(R) CPU @ 2.8 GHz and 32 GB RAM, which is a minimum node able to support a distributed simulation, and a 32 Intel(R) Xeon(R) CPU @ 2.8 GHz and 256 GB RAM, since we have accumulated up to eight nodes in the distributed simulation, both with the Debian GNU/Linux 10 operating system and OpenJDK 11.

In a first set of experiments, we have used two thread pools. The first pool was defined to run the 25\% of the slowest atomic models (49 in total), whereas the other pool was used to allocate the rest of them (149 in total, including the generator of the initial trigger event). The idea is to prove that giving resources to the slowest models (more threads, i.e., $n>m$ in Figure \ref{fig:cloud_deployment}), a better improvement in performance is obtained in return. In a second set of experiments we used a single thread pool, where the computational load of each thread was balanced allocating heaviest models in different threads. To compute the speedup, we used as the reference execution time the sequential simulation, i.e., $5896.54$ seconds. We varied the number of threads in each pool, to analyze the effects of resource allocation.

\begin{figure}
    \centering
    \begin{subfigure}{0.45\textwidth}
        \includegraphics[width=\textwidth]{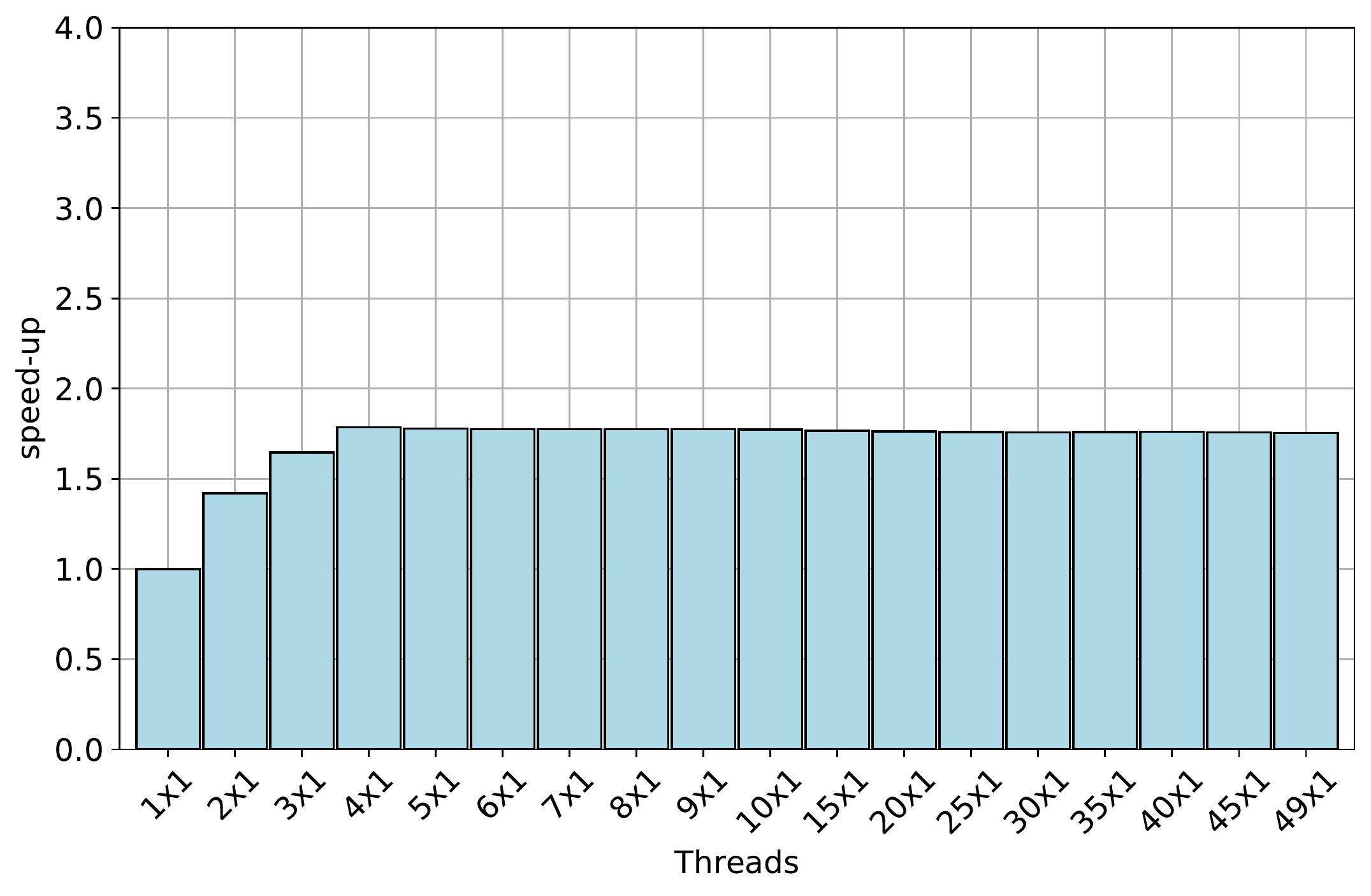}
        \caption{Varying threads for slow models}
        \label{fig:par04_spup_r1}
    \end{subfigure}\hfill
    \begin{subfigure}{0.45\textwidth}
        \includegraphics[width=\textwidth]{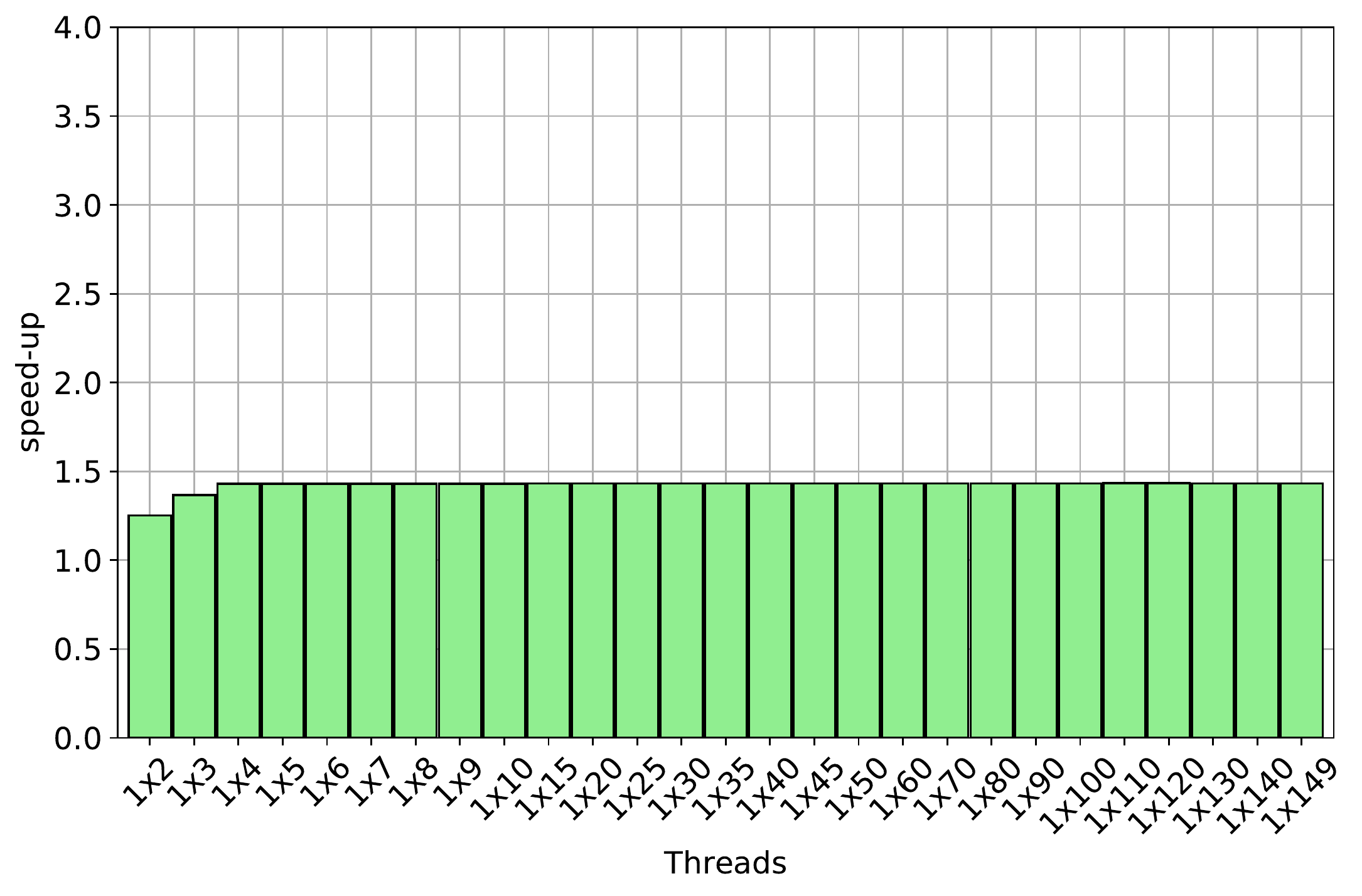}
        \caption{Varying threads for fast models}
        \label{fig:par04_spup_r2}
    \end{subfigure}
    \begin{subfigure}{0.45\textwidth}
        \includegraphics[width=\textwidth]{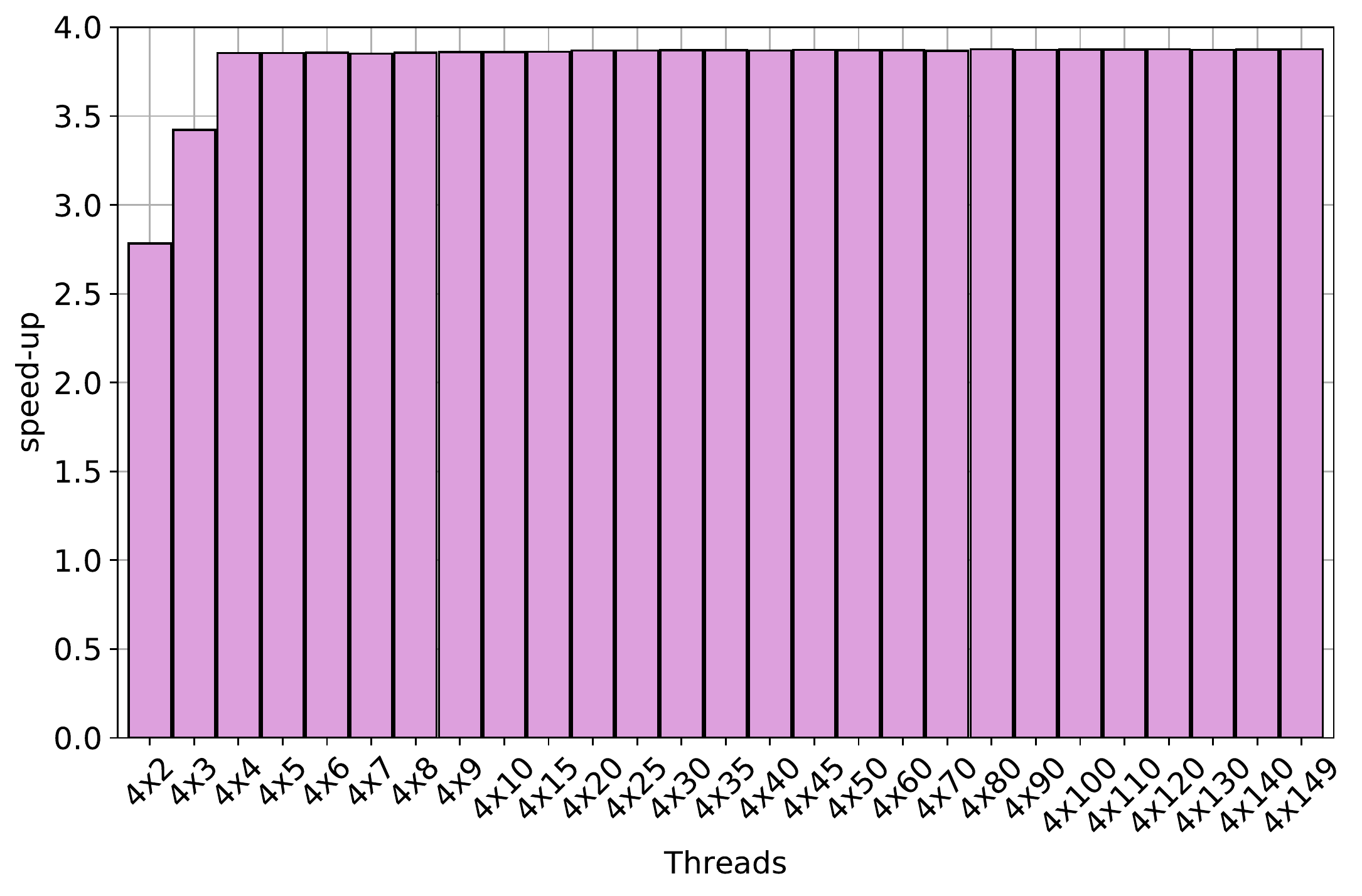}
        \caption{Sub-optimal approach}
        \label{fig:par04_spup_r3}
    \end{subfigure}\hfill
    \begin{subfigure}{0.45\textwidth}
        \includegraphics[width=\textwidth]{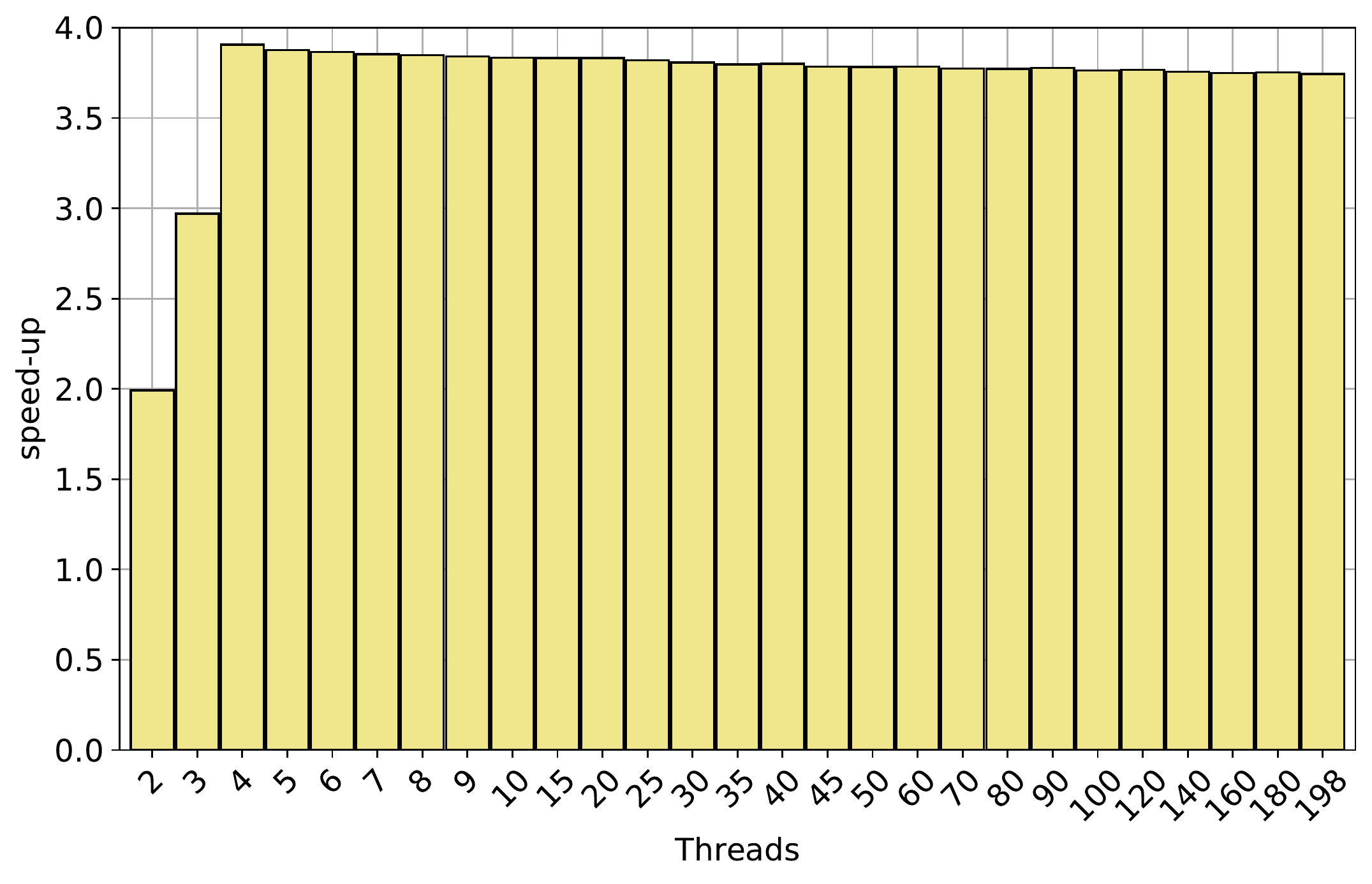}
        \caption{Balanced distribution}
        \label{fig:par04_spup_r4}
    \end{subfigure}
    \caption{Resources (threads) distribution and speedups for the 4 vCPU parallel simulation.}
    \label{fig:par04_speedup}
\end{figure}

On the one hand, Figure \ref{fig:par04_speedup} illustrate the results obtained for the 4 vCPU virtual machine. Bar labels have the form $i \times j$, where $i$ represents the number of threads in the high-priority pool ($L_1$ as in Figure \ref{fig:cloud_deployment}), and $j$ represents the number of threads in the low-priority pool ($L_2$ as in Figure \ref{fig:cloud_deployment}). Figure \ref{fig:par04_spup_r1} shows the speedup when the number of threads managed by the $L_1$ pool is being increased. As can be seen, the maximum speedup ($1.78\times$) is obtained when the $L_1$ pool uses a number of threads equal to the number of CPUs. Figure \ref{fig:par04_spup_r2} shows the same effect but varying the number of threads managed by the $L_2$ pool. However, the maximum performance in this case ($1.43\times$) is reached when the number of threads in the $L_2$ pool is equal to the number of fast atomic models (149). This is because these models have a low computational weight and then 4 CPUs are enough to handle the transition delays without difficulties. Comparing Figures \ref{fig:par04_spup_r1} and \ref{fig:par04_spup_r2}, we can observe that the speedup reached when more resources (threads) are given to the slower models is a 24.48\% greater. These two Figures confirm our $n>m$ hypothesis, since more threads for the $L_1$ pool produces a higher speed-up improvement. After that, we have looked for a sub-optimal configuration, fixing the number of optimal threads of the $L_1$ pool, $4$, and varying the number of threads in the $L_2$ pool. As Figure \ref{fig:par04_spup_r3} shows, the speedup is significantly higher ($3.88\times$ vs. the previous $1.78\times$). Finally, we ran simulations using a single thread pool, varying the number of threads. As Figure \ref{fig:par04_spup_r4} shows, the speedup obtained here is the best one ($3.91\times$). This is because when we used two thread pools, each one is executed in parallel but one pool after the other, in sequence. With one single thread pool, all the transition functions are executed in parallel, and then the linear improvement of the speedup is only limited by the number of CPUs and Input/Output operations, if any. Note that the speedup peak is always reached when the number of threads is equal to the number of CPUs. Although the forth case, with a single thread pool level, reaches the best speedup, we think based on our experience that in same real-world simulations giving resources to the slowest models can be interesting, specially when the difference between simulation times of slowest and fastest models is too high.

\begin{figure}
    \centering
    \begin{subfigure}{0.45\textwidth}
        \includegraphics[width=\textwidth]{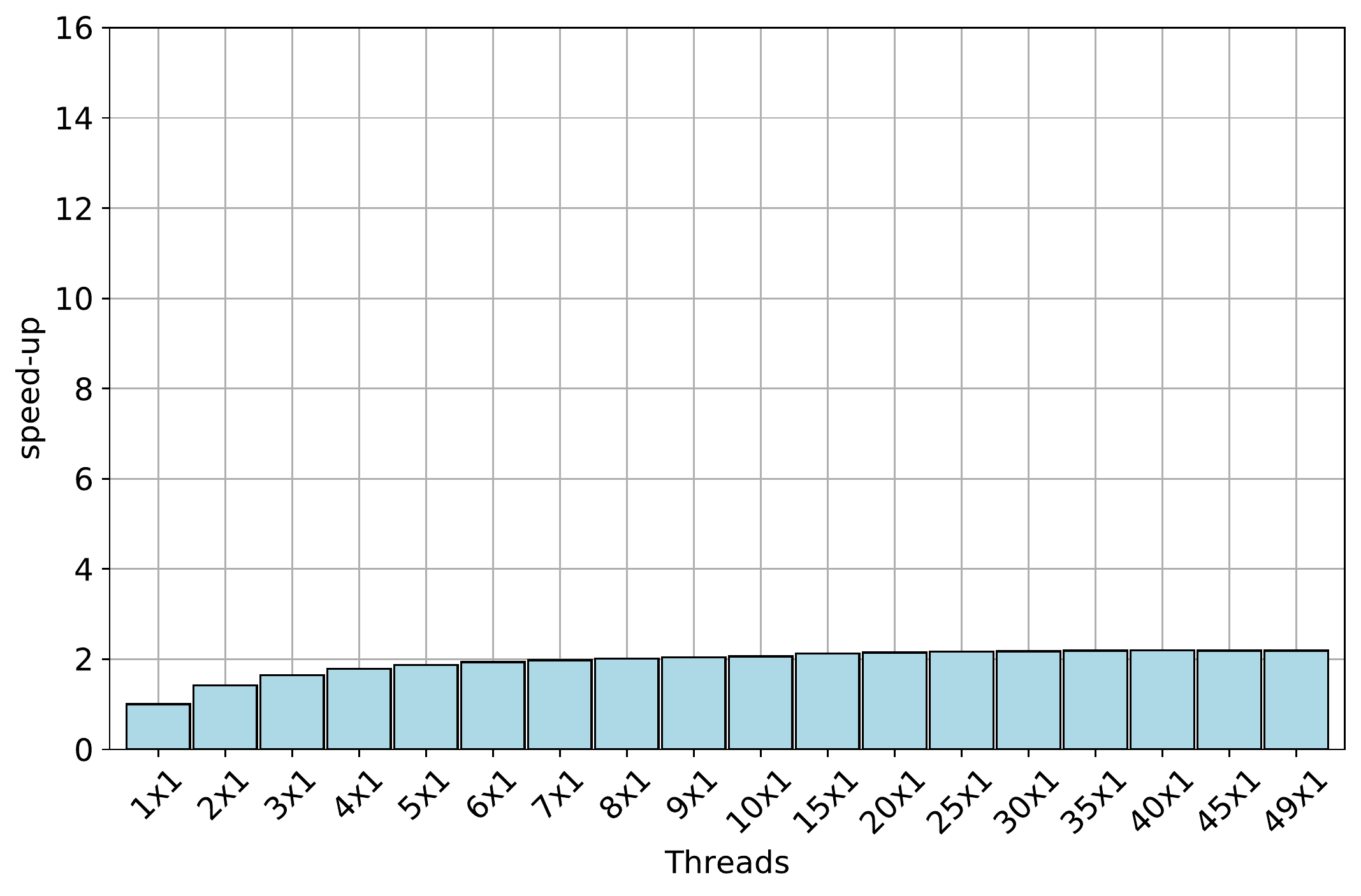}
        \caption{Threads for slow models}
        \label{fig:par32_spup_r1}
    \end{subfigure}\hfill
    \begin{subfigure}{0.45\textwidth}
        \includegraphics[width=\textwidth]{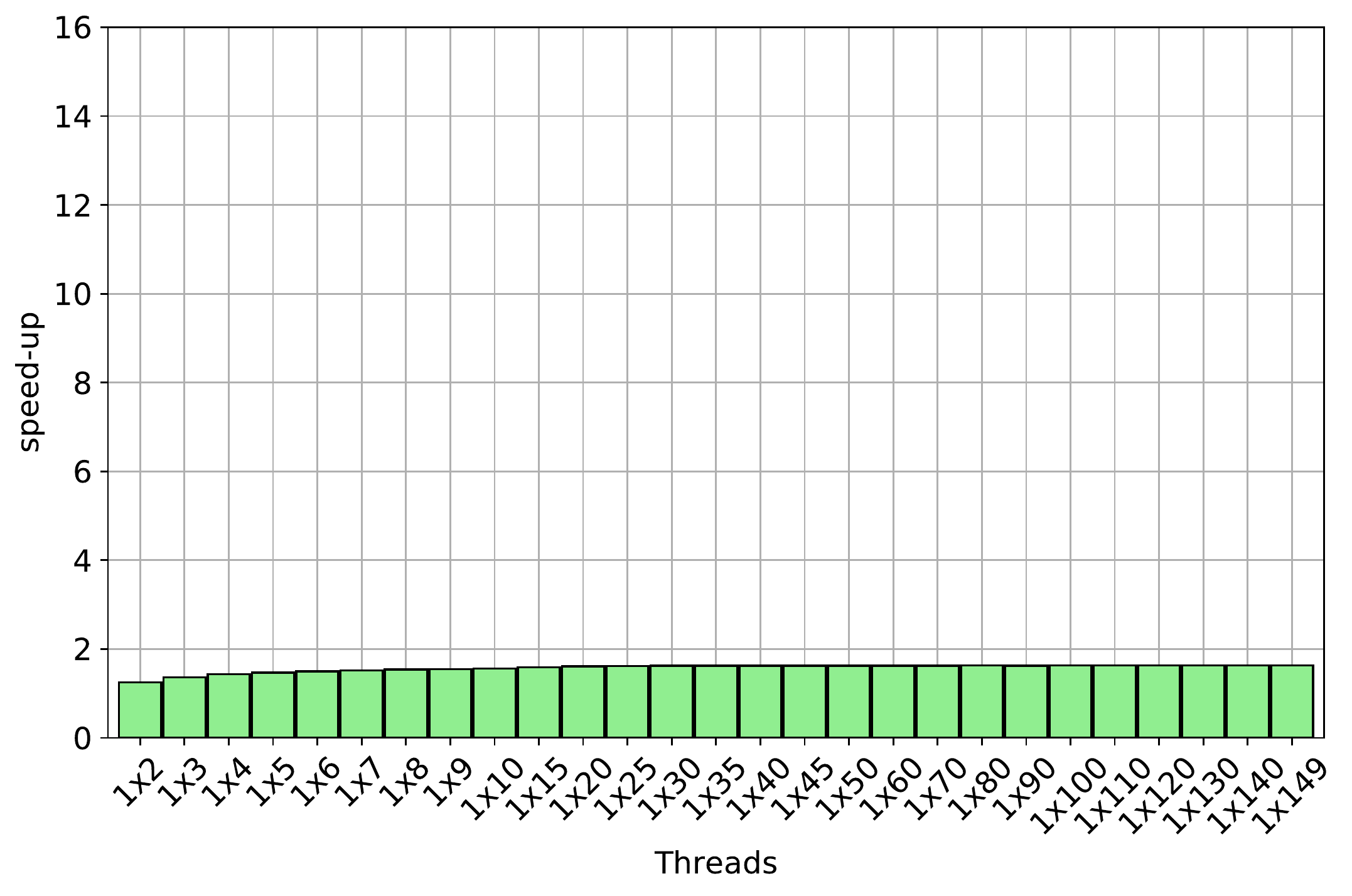}
        \caption{Threads for fast models}
        \label{fig:par32_spup_r2}
    \end{subfigure}
    \begin{subfigure}{0.45\textwidth}
        \includegraphics[width=\textwidth]{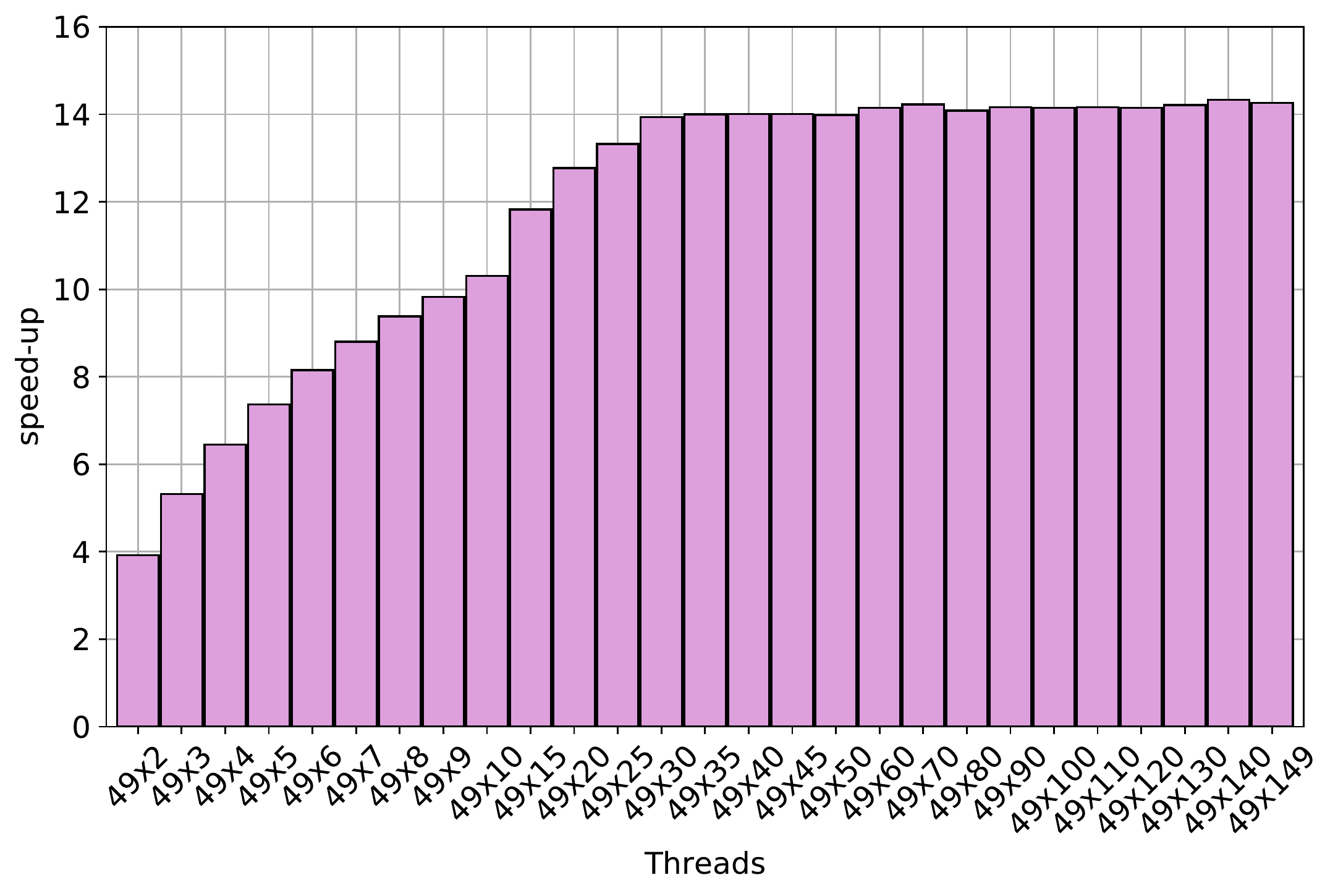}
        \caption{Sub-optimal approach}
        \label{fig:par32_spup_r3}
    \end{subfigure}\hfill
    \begin{subfigure}{0.45\textwidth}
        \includegraphics[width=\textwidth]{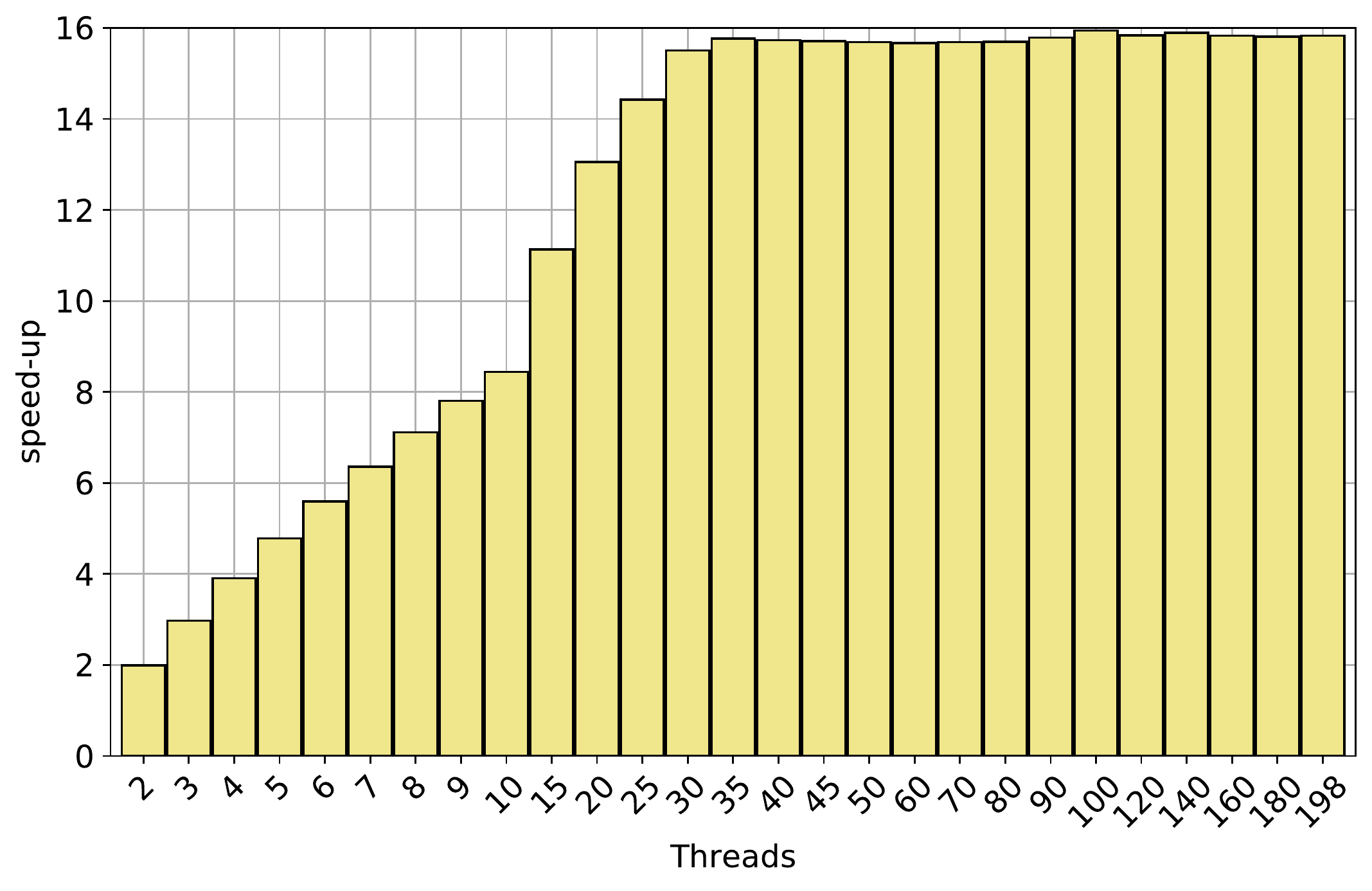}
        \caption{Balanced distribution}
        \label{fig:par32_spup_r4}
    \end{subfigure}
    \caption{Resources (threads) distribution and speedups for the 32 vCPU parallel simulation.}
    \label{fig:par32_speedup}
\end{figure}

As mentioned above, the distributed simulation used up to eight 4 vCPU 32 GB nodes. Therefore, we have repeated the previous parallel experiments on a $8 \times 4 = 32$ Intel(R) Xeon(R) CPU @ 2.8 GHz and $8 \times 32 = 256$ GiB RAM. Figure \ref{fig:par32_speedup} depicts the results. These are qualitative the same. When augmenting the number of threads in the $L_1$ pool (Figure \ref{fig:par32_spup_r1}), the maximum speedup, $2.19\times$, was obtained when the number of threads was equal to the number of slowest models: $49$, which means that 32 CPUs were able to handle all these models. The same happened when the resources went to the fast thread, i.e., the maximum speedup , $1.63\times$, was reached with a number of threads equal to the number of fast models: $149$ (Figure \ref{fig:par32_spup_r2}). As can be derived from the two previous figures as the $n>m$ option increases the more the speed-up. The sub-optimal approach (Figure \ref{fig:par32_spup_r3}) obtained a extraordinary improvement compared to the 4 CPU virtual machine, $14.33\times$ for . Additionally, as Figure \ref{fig:par32_spup_r4} illustrates, the balanced speedup in this case is much better than in the previous ones, $15.94\times$. However, there is a loss of efficiency from 4 vCPU to 32 vCPU, since $15.94 < 8 \cdot 3.91$.

\subsection{Distributed simulation}

In this section we analyze the computational cost of the distributed simulations based on the containers distribution policy and architecture described earlier in Section \ref{sec:deployment}. 

To this end, we have followed an incremental container strategy, similarly to the one used in the parallel approach, using a two-level queue (instead of thread pools) for allocation of containers (instead of threads), labeled $L_1$ and $L_2$ in Figure \ref{fig:cloud_deployment}. It is worthwhile to remind that any allocation policy can be used, editing the XML file describing the model flattened structure and the containers where each atomic is placed. As aforementioned, $L_1$ has been reserved for atomic models with high computational demands (i.e. slower atomic models), whereas level $L_2$ is used to allocate the remaining models (i.e. faster atomic models). In the first set of experiments we increased the number of containers in $L_1$ allocating one single container in $L_2$, giving more resources to the models with higher computational cost. Once we found the optimal number of container in $L_1$ (i.e. where there is no more margin for performance improvement) we then increased the number of containers in $L_2$, as we did in the parallel approach to find the 2-level sub-optimal configuration. In the second set of experiments, we simply use one single level to allocate all the containers, balancing the distribution of atomic models among them. For executing the distributed simulations, we have used a GKE cluster with 8 n2-highmem-4 nodes. These nodes count with 4 Intel(R) Xeon(R) vCPU @ 2.8 GHz and 32 GiB of RAM.

\begin{figure}
\centering
\includegraphics[width=1\textwidth]{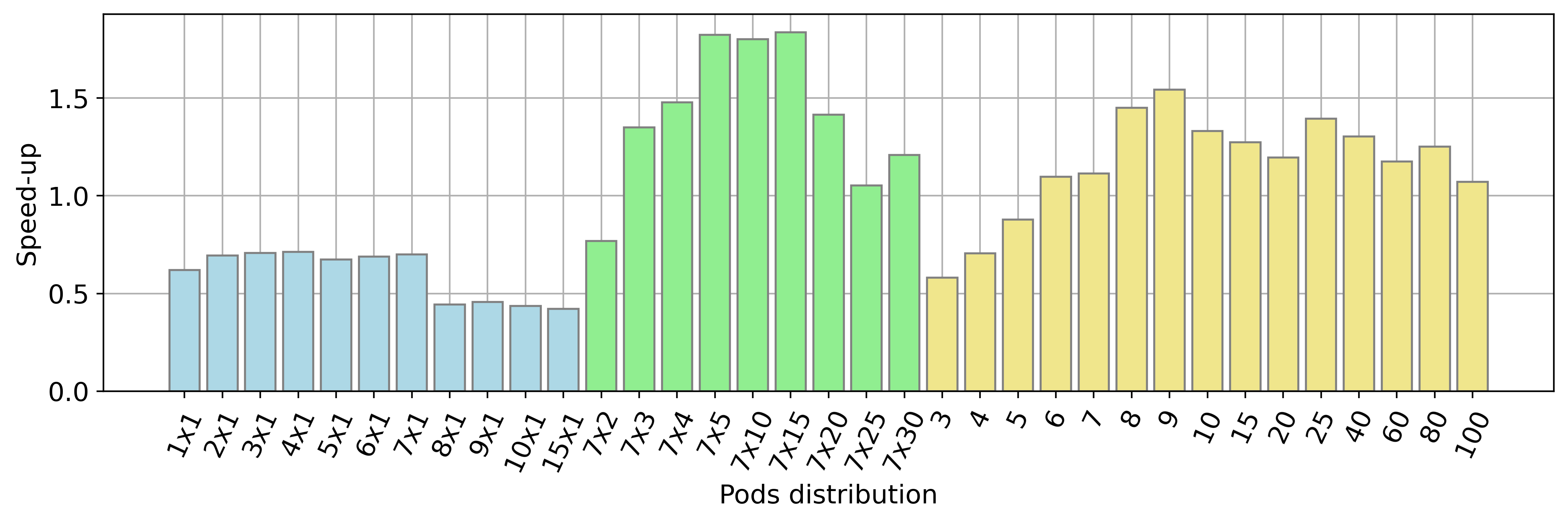}
\caption{Distributed simulations speed-ups depending on the number of pods ($L_1 \times L_2$).}
\label{fig:distributed_sim_times}
\end{figure}

Figure~\ref{fig:distributed_sim_times} depicts the results of this analysis in terms of speed-up. As in the parallel simulations, bars are labeled as $i \times j$, where $i$ represents the number of containers (pods) created in $L_1$ and $j$ pods created in $L_2$. As we can see in the blue bars of Figure~\ref{fig:distributed_sim_times}, the speed-up is increased as the number of pods created for slower models in $L_1$ increases, reaching a maximum value of $0.71\times$ in $7\times1$. However, the results suddenly become worse starting from the $8\times1$ distribution in advance. This is because of the number of nodes present in the cluster. While the $7\times1$ scenario distributes exactly one pod per node, the following scenarios present nodes with multiple pods. The speed-up is less than 1 in this case. This is because the distributed simulation differs from the parallel one mainly in which all the 198 simulators are executed as independent Java Virtual Machine (JVM) processes, independently of the number of pods, i.e., memory resources needed to run the distributed simulation is significantly higher than in the parallel solution. As in the $8\times1$ configuration there are more containers than nodes, there are also less resources for the execution of the 149 fastest models, abruptly increasing the execution time and decreasing the speed-up in consequence. As a result, the optimal number of containers in $L_1$ with a single container in $L_2$ is reached when there are 7 containers in $L_1$. Beyond this number, there is no benefit in increasing the number of containers in $L_1$ without increasing the number of containers in $L_2$. 

After that, the number of containers in $L_1$ is fixed and equal to the optimal value and then the $L_2$ size in increased, looking for a sub-optimal configuration as in the parallel case. This can be seen in the green bars of Figure~\ref{fig:distributed_sim_times}, which show that increasing containers in $L_2$ also improves the performance notably, with a maximum speedup value of $1.84\times$ at $7\times15$. This fact does not reinforce our $n>m$ hypothesis, because the 149 Java Virtual Machine (JVM) instances consume significant amounts of memory and becomes a bottleneck, favoring a higher value for $m$, which also explains the poor speed-up value.

Finally, we have used a single container level, i.e., one single class to allocate all the atomic models, balanced according to their delays. The yellow bars in Figure \ref{fig:distributed_sim_times} shows that this configuration can give up to $1.45\times$. Again, the speed-up increases with the number of pods, until those are approximately equal to the number of nodes. In the distributed version, the benefits of dividing the resources in levels is not as clear as in the parallel version, since as stated above (a) all the atomic models are executed concurrently, and (b) one of the levels can act as bottleneck when the resources (mainly memory) reserved for that level are insufficient. This inefficiency might be attributed to propagation issues as is the case in distributed simulations. However, we confirmed that the bottleneck is not the communication between nodes, because setting the delays equal to 0 seconds ($\Delta_i=0$), the speed-ups obtained by the 32 CPU parallel machine and the distributed version were equivalent ($21\times$ vs. $19\times$). Future work includes the study of mechanisms to alleviate the weight of the JVM processes.

\subsection{Parallel vs. Distributed}

In order to compare our parallel and distributed architecture, four metrics must be considered: performance, cost, cost/performance, and underlying hardware. Table \ref{tab:comparison} shows the best speedup obtained by the suboptimal and balanced configurations and the monthly cost of the nodes used for the parallel and distributed simulations.

\begin{table}
    \centering
    \begin{tabular}{|c|c|c|c|}
          & \textbf{Sub-optimal} & \textbf{Balanced} & \textbf{\$/month} \\
        \hline
        Parallel 4 vCPU	                & 3.88	        & 3.91  & 168.38 \\
        Parallel 32 vCPU                & 14.33	        & 15.94 & 1413.63\\
        Distributed 8 $\times$ 4 vCPU   & 1.84    & 1.45 & 1347.01\\
        \hline
    \end{tabular}
    \caption{Maximum speedup and monthly cost of the parallel and distributed simulations.}
    \label{tab:comparison}
\end{table}

As Table \ref{tab:comparison} shows, the best performance is obtained with the Parallel 32 vCPU balanced configuration, nearly 16 times faster than the sequential simulation. 

With respect to cost, the cheapest solution is of course the 4 vCPU parallel approach, with a monthly cost of \$168, as can be seen in Table \ref{tab:comparison}. It is followed by the Kubernetes cluster, with \$1347/month. Finally, the 32 vCPU virtual machine is the most expensive solution with \$1414/month. The cost/perfomance of a single balanced speedup point is \$43, \$89 and \$732 for the 4, 32 vCPU VMs and distributed solutions, respectively. Clearly, the distributed infrastructure is completely saturate and must not be pursued from a cost-benefit factor.

Finally, regarding underlying hardware, the distributed solutions is more flexible, since it supports heterogeneous architectures as the simulation is based on a socket distributed application, compatible with any hardware distribution. The parallel solutions is only valid for systems with shared memory and homogeneous architecture.

There is still much work to do in the field of distributed simulations. Obviously, memory management by independent distributed processes is a huge bottleneck that must be alleviated. In any case, regarding the possible difficulties around the distributed setup, the M\&S framework presented in this paper allows us a unified sequential, parallel, and distributed solution that facilitates the deployment of any configuration, being completely focused on model immutability and automated deployment.

\section{Conclusions and future work}\label{sec:conclusion}
Simulation is an activity of running a simulator in a computational environment. This computational environment has evolved over time. Today it consists of varied options such as local desktop, distributed network, multi-core, virtualized infrastructure, HPC infrastructure and cloud-enabled containerized environment. An extensive and scalable simulation architecture must be able to execute in any computational environment in a seamless manner. Unfortunately, most simulation architectures are not designed to be extensible and scalable, especially when a large number of simulation runs are needed from a model that was designed for a local desktop execution that is unable to run in other high performance environments. It is a well known fact that sequential programs that were designed for a single CPU receive no benefit from their execution on a multi-core CPU. The same is true for simulation architectures. The problem is more compounded when the model formalism is tightly coupled with the simulation architecture and both need to be rewritten for execution in a different computational execution environment than the original. 

DEVS formalism categorically separates the modeling and simulation layers so that the simulation architecture is transparent to the model architecture and both can evolve horizontally. Over the past 15 years, the work by Mittal and Martin have demonstrated this aspect of executing DEVS models and various Domain Specific Models (DSMs) with their DEVS mappings over transparent simulation architectures . This paper has provided evidence that advances their earlier work with the xDEVS M\&S simulation engine capable of deploying the simulator in a parallel multi-core architecture and in a distributed networked architecture in a seamless manner. We have described a unifying architecture  incorporating two DEVS coordinators that run the same DEVS model in both parallel and distributed architectures. While this basic concept of having different coordinators for different deployment platforms was introduced in Zeigler's text \cite{Zeigler2000}, it needed some improvements for their usage in cloud-enabled platforms. These two coordinators were further deployed in an cloud-enabled containerized environment making the simulation infrastructure truly transparent to the model. Both parallel and distributed implementations are DEVS-compliant. This assures that the sequential, parallel and distributed simulations provide exactly the same results.

We described the performance evaluation of the Parallel simulation coordinator and the Distributed simulation coordinator using the DEVStone benchmark and conclusively received a 16 times speedup by the Parallel coordinator and 1.84 times speedup by the Distributed coordinator for a given hardware configuration. For Parallel simulation, we achieved the following:
\begin{enumerate}
    \item Confirmed our hypothesis that more thread assignment to the thread pool that contains CPU-intensive models produces a higher speedup improvement.
    \item Speedup peak is achieved when the number of threads in a thread pool is equal to the number of CPUs.
    \item Cost-benefit factor is much higher as compared to distributed simulation performance use case
\end{enumerate} 

This result demonstrates that the parallel execution of any DEVS model must be preferred over any distributed execution. This result is further extended to the entire case of distributed simulation that can never match the results obtained by parallel architectures. 

The distributed computing architectures came before the multi-core parallel computing architectures. The motivation for distributed computing (before the ubiquitous Internet), which was to connect geographically distributed entities to solve a complex problem has now given  way to parallel computing wherein the resources are made available either in HPC or Cloud-environments and are transparently available for use. Accordingly, the M\&S architectures (both legacy and upcoming) must evolve to benefit from the cloud-enabled parallel computing architectures. Adhering to formalisms such as DEVS (and the associated xDEVS implementations) that provide sound basis for composable M\&S architectures is the preferred way to go. Various algorithms, features and APIs developed in xDEVS framework provide ease of use, extensibility and scalability to any DEVS simulation. xDEVS has been reported as the most efficient DEVS simulator till date \cite{J-RiscoMartin2017} and this work extends its capability to cloud-enabled parallel and distributed simulation.

While the parallel simulation architectures provide speedup to run simulations in a high performance environment for running optimizations and analyses on the model, the distributed simulation architectures will continue to find their niche in training, testing and evaluation of interoperability in Live, Virtual and Constructive environments and integration of new systems for human-in-the-loop experimentation. 

\subsection{Future Work}
We established the case for the increased usage of parallel architecture as compared to distributed architectures. However, there is value to be had in bringing these two architectures together for maximum value. Future work includes the exploitation of such hybrid deployments, where distributed nodes can perform parallel simulations. This would require modifying the xDEVS modeling layer and demands a major development effort. We are also considering the a comparative study with other DEVS simulation engines, when they incorporate a parallel interface and a supporting unifying architecture. Finally, though we have artificially added CPU stress to focus on computing performance, data exchange and network latency analysis is also of great interest.

\section*{Acknowledgments}
This project has been partially supported by the Education and Research Council of the Community of Madrid (Spain), under research grant S2018/TCS-4423, and by the Google Cloud Research Credits program with the award GCP19980904.

\section*{Disclaimer}
The author’s affiliation with The MITRE Corporation is provided for identification purposes only, and is not intended to convey or imply MITRE’s concurrence with, or support for, the positions, opinions or viewpoints expressed by the author(s). ©2021 The MITRE Corporation. ALL RIGHTS RESERVED. Approved for Public Release. Distribution Unlimited. Case Number 21-02817-1.

\bibliographystyle{elsarticle-num}
\bibliography{biblio}

\end{document}